\documentclass[conference]{IEEEtran}
\IEEEoverridecommandlockouts

\usepackage{cite}
\usepackage{amsmath,amssymb,amsfonts}
\usepackage{algorithmic}
\usepackage{graphicx}
\usepackage{textcomp}
\usepackage{xcolor}
\usepackage{cleveref}
\usepackage{booktabs}
\usepackage{breqn}
\usepackage{url}

\def\BibTeX{{\rm B\kern-.05em{\sc i\kern-.025em b}\kern-.08em
    T\kern-.1667em\lower.7ex\hbox{E}\kern-.125emX}}

\begin{document}

\title{Efficient Multiparty Protocols Using Generalized Parseval’s Identity and the Theta Algebra}

\author{\IEEEauthorblockN{Giorgio Sonnino}
  \IEEEauthorblockA{
    \textit{Université Libre de Bruxelles (ULB)}\\
    Brussels, Belgium \\
    giorgio.sonnino@ulb.be}
  \and
  \IEEEauthorblockN{Alberto Sonnino}
  \IEEEauthorblockA{
    \textit{Mysten Labs}\\
    London, U.K. \\
    alberto@mystenlabs.com
  }
}
\maketitle

Manuscript Accepted for Publication in {\it International Conference on Mathematics and Computers in Sciences and Industry} by the IEEE {\it Computer Society Conference Publishing Services}.
\vskip0.5truecm
\begin{abstract}
We propose a protocol able to show publicly addition {\it and} multiplication on secretly shared values. To this aim, we developed a protocol based on the use of masks and FMPC (Fourier Multi-Party Computation). FMPC is a novel multiparty computation protocol of arithmetic circuits based on secret-sharing, capable to compute the addition and multiplication of secrets with no communication. We achieve this task by introducing the first generalization of Parseval’s identity for Fourier series applicable to an arbitrary number of inputs and a new algebra referred to as the $\Theta^{[n]}$-{\it algebra}. FMPC operates in a setting where users wish to compute a function over some secret inputs by submitting the computation to a set of nodes, without revealing those inputs. FMPC offloads most of the computational complexity to the end users and includes an online phase that mainly consists of each node locally evaluating specific functions. FMPC paves the way for a new kind of multiparty computation protocol; making it possible to compute the addition and multiplication of secrets stepping away from circuit garbling and the traditional algebra introduced by Donald Beaver in 1991. Our protocol is capable to compute addition and multiplication with no communication and its simplicity provides efficiency and ease of implementation.

\end{abstract}


\section{Introduction}\label{I}
MPC (Multi-Party Computations) are cryptographic protocols where several distinct, yet connected, computing devices (or parties) jointly evaluate a public function while preserving several security properties despite adversarial behavior~\cite{SPDZ}. This work aims to solve the following problem: {\it Develops a method able to show publicly the result of a general mathematical expression while keeping the inputs of the expression secret}. This problem must be solved by satisfying the following conditions:
\vskip0.2truecm
\noindent {\bf 1)} Users are indistinguishable from each other;

\noindent {\bf 2)} Users cannot communicate with each other;

\noindent {\bf 3)} All operations must be performed simultaneously for all users;

\noindent {\bf 4)} The number of nodes that are not corrupted must not depend on the number of nodes involved in the process;

\begin{figure*}[t]
  \centering\includegraphics[scale=.15]{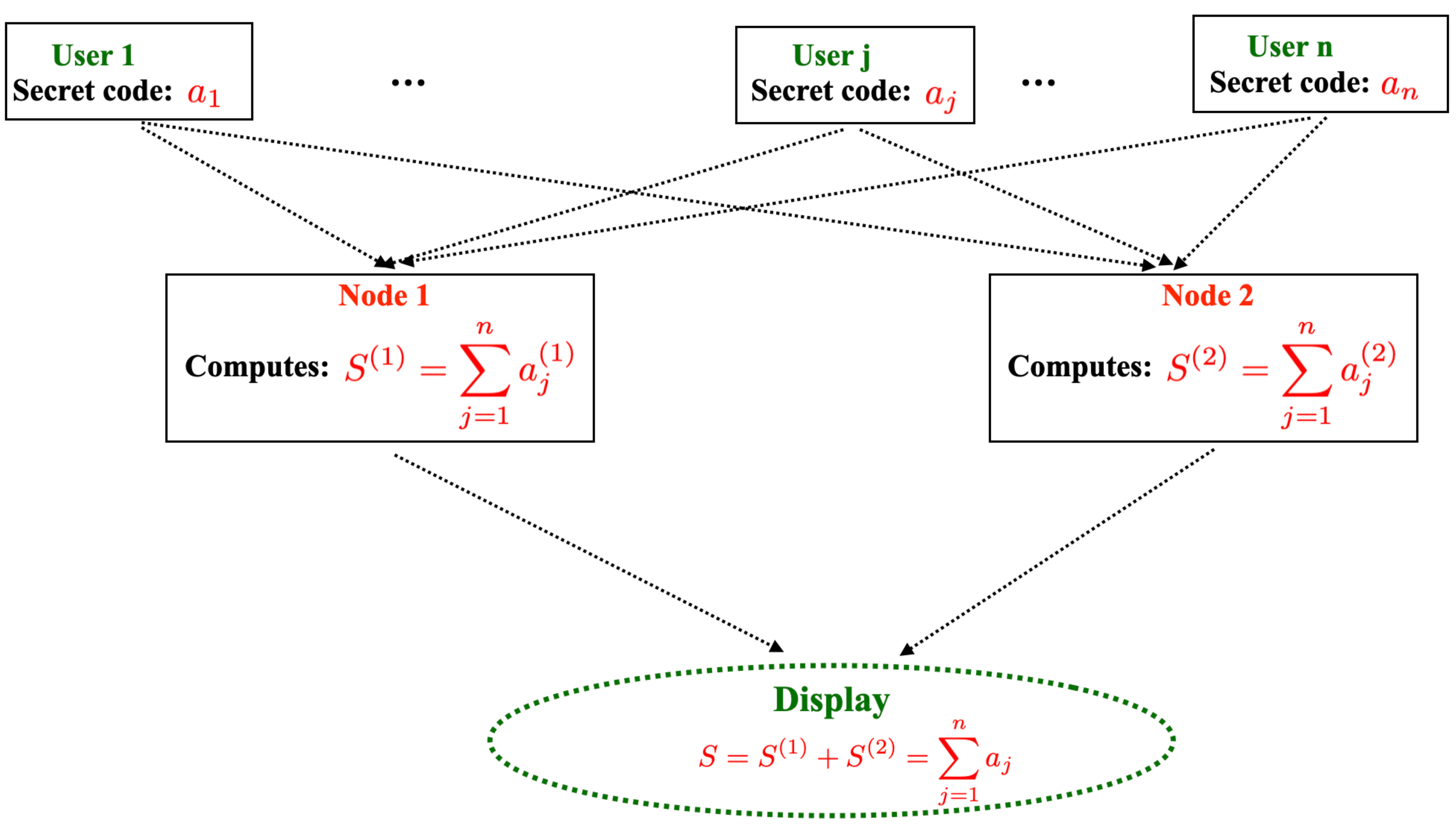}
  \caption{{\bf Display of the expression $S=\sum_{j=1}^na_j$ by keeping secret the codes $a_j$.} This scheme illustrates how an expression, composed of a sum of secret codes only, can be shown publicly by keeping secret the codes of single users. The users split their code in two pieces $a_j^{(1)}$ and $a_j^{(2)}$. The values of $a_j^{(1)}$ and $a_j^{(2)}$ may be chosen such that $a_j=a_j^{(1)}a_j^{(2)}$ or, by using additive masks $\omega_j$, $a_j^{(1)}\equiv a_j+\omega_j$ and $a_j^{(2)}\equiv -\omega_j$. At least, one node is not corrupted.}
  \label{fig_sum}
\end{figure*}

\noindent {\bf 5)} The operations performed by the display are visible to the public.
\vskip0.2truecm
\noindent It is trivial to solve the problem when the mathematical expression is composed only of sums or only by multiplications of (secret) codes. Indeed, in the first case (i.e., the expression is made up of sums of the codes only) it is sufficient that each user splits the codes into two contributions and sends them to two distinct nodes that cannot communicate with each other, according to the following procedure. Denoting with $a_j$ the secret code of the user $j$, with $j=1,\cdots, n$, each user splits its code in two parts, $a_j^{(1)}$ and $a_j^{(2)}$, such that $a_j=a_j^{(1)}+a_j^{(2)}\ \ \forall j$. Successively, they send the contribution $a_j^{(1)}$ to the $Node_1$ and the contribution $a_j^{(2)}$ to the node $Node_2$, respectively. $Node_1$ and $Node_2$ perform the partial sum $S^{(1)}=\sum_{j=1}^na_j^{(1)}$ and $S^{(2)}=\sum_{j=1}^na_j^{(2)}$, respectively. Finally, the {\it Display} shows publicly the value of the sum $S$ with $S=S^{(1)}+S^{(2)}$ (see Figure~\ref{fig_sum}).
\noindent Similarly, when the expression is made up of products of the codes only, it is sufficient that each user splits its code in two parts, $a_j^{(1)}$ and $a_j^{(2)}$, such that $a_j=a_j^{(1)}\cdot a_j^{(2)}\ \ \forall j$. Successively, they send the contribution $a_j^{(1)}$ to the $Node_1$ and the contribution $a_j^{(2)}$ to the node $Node_2$, respectively. $Node_1$ and $Node_2$ perform the partial product $P^{(1)}=\Pi_{j=1}^na_j^{(1)}$ and $P^{(2)}=\Pi_{j=1}^na_j^{(2)}$, respectively. Finally, the {\it Display} shows publicly the value of the product $P$ with $P=P^{(1)}\cdot P^{(2)}$ (see Figure~\ref{fig_product}):
\begin{figure*}[t]
  \centering\includegraphics[scale=.15]{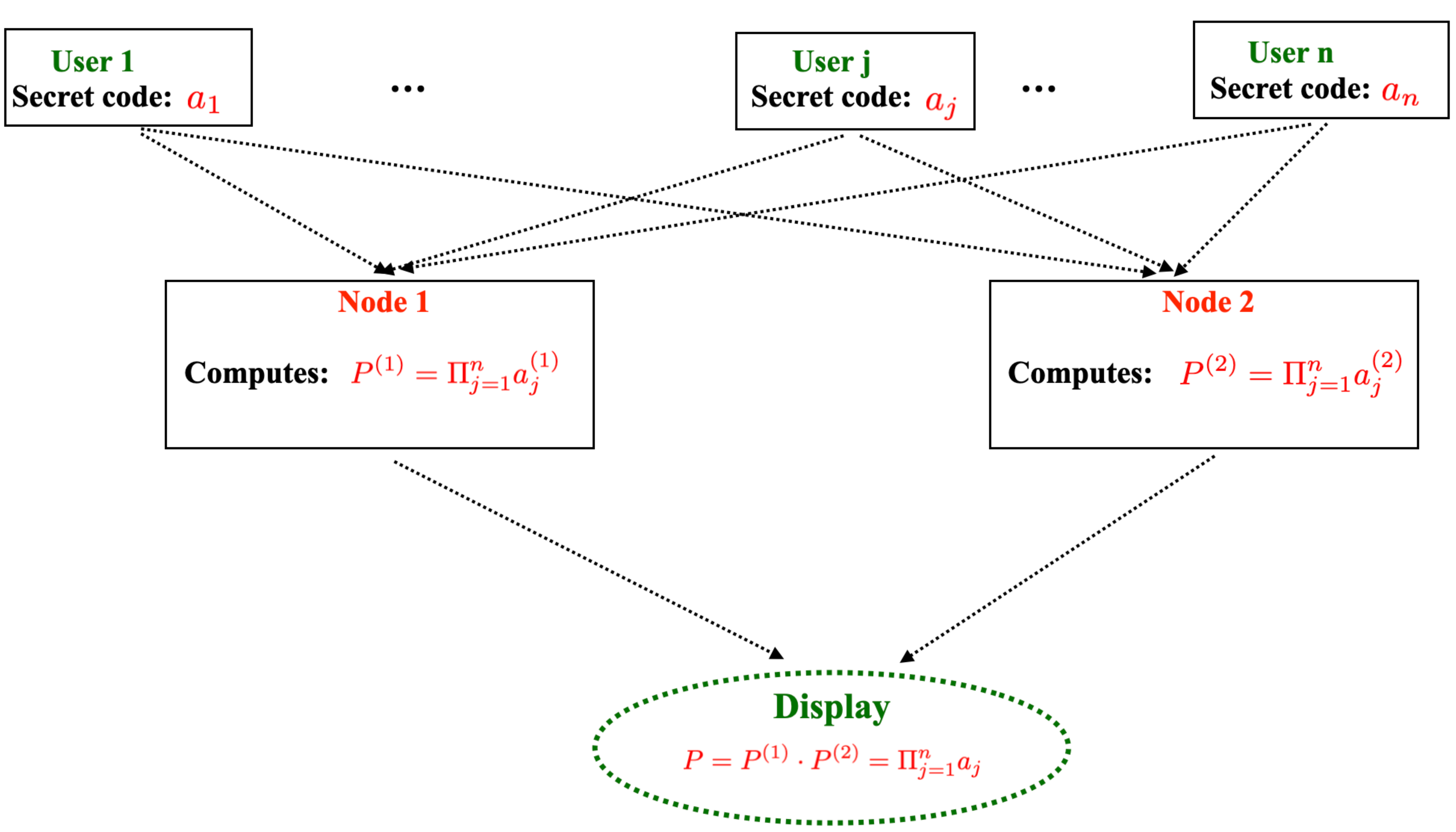}
  \caption{{\bf Display of the expression $P=\Pi_{j=1}^na_j$ by keeping secret the codes $a_j$.} This scheme illustrates how an expression composed only by a product of secret codes can be shown publicly by keeping secret the codes of single users. The users split their code in two pieces $a_j^{(1)}$ and $a_j^{(2)}$. The values of $a_j^{(1)}$ and $a_j^{(2)}$ may be chosen such that $a_j=a_j^{(1)}a_j^{(2)}$ or, by using multiplicative masks $\omega_j$, $a_j^{(1))}\equiv |a_j|\omega_j$ and $a_j^{(2))}\equiv {\widetilde\omega}_j$ with ${\widetilde\omega}_j$ defined such that $a_j=|a_j|\omega_j{\widetilde\omega}_j$.At least one node is not corrupted. }
  \label{fig_product}
\end{figure*}

\noindent Another very efficient way to solve the previous problem is based on the use of numerical {\it masks}.

\noindent As for the sum of the codes, the users chose {\it additive masks} $\omega_j$ ($j=1,\cdots, n$) and split the code in two parts: $a_j^{(1)}=a_j+\omega_j$ and $a_j^{(2)}=-\omega_j$ and they send $a_j^{(1)}$ and $a_j^{(2)}$ to the the $Node_1$ and $Node_2$, respectively. Successively, $Node_1$ and $Node_2$ perform the partial sums  $S^{(1)}=\sum_{j=1}^na_j^{(1)}$ and $S^{(2)}=\sum_{j=1}^na_j^{(2)}$, respectively. Finally, the {\it Display} shows publicly the value of the sum $S$ with $S=S^{(1)}+S^{(2)}$.

\noindent When the expression is made only by a multiplication of the codes, the users chose the {\it multiplicative masks} $\omega_j$ and ${\widetilde\omega}_j$ ($j=1,\cdots, n$) such that $|a_j|\omega_j{\widetilde\omega}_j=a_j$ with $|a_j|$ denoting the absolute value (i.e., the modulus) of the code $a_j$\footnote{Notice that by choosing the masks $\omega_j$ in such a way that $|a_j|\omega_j{\widetilde\omega}_j=a_j$, it is not possible to determine neither the  value nor the sign of the code $a_j$}. Successively, the users send the values of $|a_j|\omega_j$ and ${\widetilde\omega}_j$ to $Node_1$ and $Node_2$, respectively. $Node_1$ and $Node_2$ perform the partial product $P^{(1)}=\Pi_{j=1}^n|a_j|\omega_j$ and $P^{(2)}=\Pi_{j=1}^n{\widetilde\omega}_j$, respectively. Finally, the {\it Display} shows publicly the numerical value of the product $P$ with $P=P^{(1)}\cdot P^{(2)}$.

\noindent Another important problem that we have to solve is the determination of a procedure able to use, instead of two nodes, an arbitrary number of nodes $N$, {\it while keeping constant the number of nodes that must not be corrupted}. It is customary to enumerate the nodes with $N=3f+1$ with $f$ denoting a natural number (i.e., $f=1,2,\cdots$). It is easy to set up this procedure when the mathematical expression is made only by sums of the codes or products of the codes. Indeed, it is sufficient for each user to divide their code into $2f+1$ pieces:
\begin{eqnarray}\label{I1}
  &&a_j=\sum_{i=1}^{3f+1}a_j^{(i)}\quad\ \ {\rm in\ case\ of\ sums}\nonumber\\
  &&a_j=\Pi_{i=1}^{3f+1}a_j^{(i)}\quad{\rm in\ case\ of\ products}\nonumber
\end{eqnarray}
\noindent and then send each piece to a different node. The numerical value of the expression can be shown publicly without revealing the secrets provided that at least one node is not corrupted. Of course, the same procedure applies when using {\it masks}.

\noindent To sum up when the expression contains either sums or products of the (secret) codes, the problem of showing publicly the numerical value of this expression is trivially solved. The issue arises when the expression is composed of a combination of sums and products of codes. This work aims to present a protocol able to show publicly addition {\it and} multiplication on secretly shared values. This problem has already been solved by Donald Beaver in 1991 by using traditional algebra - essentially based on the number theory - and a wide series of circuits garbling \cite{beaver1991efficient}. Unfortunately, Beaver's protocol requires many rounds of communication and this slows down the machine time with consequent expensive energy. To overcome these obstacles we resort to a more sophisticated mathematical approach, based on the use of masks and FMPC (Fourier-based Multi-Party Computation)~\cite{sonnino2019fmpc}. Our protocol operates by using the masks’ method and paves the way for a new kind of multiparty computation protocol capable of computing sums and multiplications of secrets as an alternative to circuit garbling. More precisely, we hide the users' input codes within the cosine components of the Fourier series of the {\it main function} (denoted by $f(x)$), combined with {\it additive masks} that can be chosen arbitrarily by the users. The result of the function is visible publicly but, of course, the user's mask is kept secret. Users send the secret codes hidden by the masks to four nodes. Computations are performed by using a new algebra, referred to as the $\Theta^{[n]}$-{\it algebra} (or {\it Theta-algebra}). Finally, the results of the calculation are transmitted to the display which, thanks to the {\it generalized Parseval's identity for Fourier series} applicable to an arbitrary number of inputs, can show publicly the numerical result of the mathematical expression while keeping secret the users' codes. The participants' privacy is guaranteed if at least three nodes are not corrupted. Successively, we solved the problem using $3f+1$ nodes, grouped in four {\it categories}. In this case users, besides the additive mask, arbitrarily choose another {\it multiplicative mask}. Even for multiple nodes, the participants' privacy is guaranteed if at least three nodes belonging to three different categories are not corrupted and at least one node, belonging to the {\it second level} of computation, is not corrupted. Compared to the original Beaver's method, our protocol is capable to compute addition and multiplication with {\it no online communication} and its simplicity provides efficiency and ease of implementation.

\noindent The manuscript is organized as follows.  Section~\ref{secret2Dim} presents a protocol capable to show publicly addition {\it and} multiplication for two players, This task is accomplished by using additive masks and FMPC. Section~\ref{GE} is devoted to the application of the theorems valid for the Chebyshev polynomials. We shall see that these theorems allow showing a general mathematical expression while keeping secret the users' codes. The mathematical framework to be used for treating the case of $n$ players is described in Section~\ref{math}. In Section \ref{nU} we treat the problem for $n$ players by using the generalized Parseval's identity and the  $\Theta^{[n]}$-{\it algebra}. Finally, in Section~\ref{nUN} we solve the problem for $3f+1$ nodes, grouped in four {\it categories}. In this case, users need to choose, arbitrarily, multiplicative as well as additive masks. The generalized Parseval's identity and the use of the {\it Theta-algebra} ensure the correct result shown by the display. Section~\ref{cons} presents designs that incorporate our nodes within the infrastructure of several semi-permissioned Blockchains. Limitations of our work and concluding remarks can be found in Section~\ref{sec:limitations} and Section~\ref{conc}, respectively. Useful relations for getting the values of the infinite sums, the proof of the generalized Parseval's identity, and the rules for the $\Theta^{[n]}$-{\it algebra} are reported in Appendices~\ref{A}, \ref{B}, and \ref{C}, respectively.

\section{Display of the expression $x_1a+x_2b+yab$ by keeping secret the codes $a$ and $b$.}\label{secret2Dim}

\subsection{Background}
\noindent We recall the Fourier series of the convolution between two functions $\phi(x)$ and $\psi(x)$ periodic on ($-l,\ l$). Assuming that $\phi(x)$ and $\psi(x)$ $\in{\mathbb L}^2[-l,l]$ ($\phi(x)$ and $\psi(x)$ are square-integrable in the interval [$-l,\ l$]), their respective Fourier series representations read:
\begin{eqnarray}\label{conv1}
  &&  \phi(x)=\frac{\alpha^{(0)}}{2}+\sum_{m=1}^{+\infty}\alpha_{m}\cos\Bigl(\frac{n\pi x}{l}\Bigr)+\sum_{m=1}^{+\infty}\beta_{m}\sin\Bigl(\frac{n\pi x}{l}\Bigr)\nonumber\\
  &&  \psi(x)=\frac{a^{(0)}}{2}+\sum_{m=1}^{+\infty}a_{m}\cos\Bigl(\frac{n\pi x}{l}\Bigr)+\sum_{m=1}^{+\infty}b_{m}\sin\Bigl(\frac{n\pi x}{l}\Bigr)\nonumber
\end{eqnarray}
where the Fourier coefficients $(\alpha^{(0)},\ \alpha_{m},\ \beta_{m})$ and $(a^{(0)},\ a_{m},\ b_{m})$ (for $m=1,2,\cdots)$ are given below:
\begin{align}\label{conv2}
  \alpha^{(0)} & =\frac{1}{l}\int_{-l}^{l}\phi(x)dx \; ; \quad  \alpha_{m}=\frac{1}{l}\int_{-l}^{l}\phi(x)\cos\Bigl(\frac{m\pi x}{l}\Bigr)dx \nonumber \\
  \beta_{m}    & =\frac{1}{l}\int_{-l}^{l}\phi(x)\sin\Bigl(\frac{m\pi x}{l}\Bigr)dx \nonumber                                                          \\
  a^{(0)}      & =\frac{1}{l}\int_{-l}^{l}\psi(x)dx \; ; \quad a_m=\frac{1}{l}\int_{-l}^{l}\psi(x)\cos\Bigl(\frac{m\pi x}{l}\Bigr)dx \nonumber         \\
  b_{m}        & =\frac{1}{l}\int_{-l}^{l}\psi(x)\sin\Bigl(\frac{m\pi x}{l}\Bigr)dx\nonumber
\end{align}
\noindent Parseval's identity holds for $f(x)$ and $g(x)$ \cite{gradshteyn2014table}:
\begin{equation}\label{pars1}
  \Big(\frac{a_0\alpha_0}{2}+\sum_{m=1}^\infty a_m\alpha_m\Big)+\Big(\sum_{m=1}^\infty b_m\beta_m\Big)=\frac{1}{l}\int_{-l}^l \phi(x)\psi(x)dx
\end{equation}
\noindent We introduce the {\it normalised functions} $f(x)=\eta^{-1/2}\phi(x)$ and  $g(x)=\eta^{-1/2}\psi(x)$ with
\begin{equation}\label{pars2}
  \eta=\frac{1}{l}\int_{-l}^l \phi(x)\psi(x)dx
\end{equation}
\noindent For functions $f(x)$ and $g(x)$, the Parseval identity reads:
\begin{equation}\label{pars3}
  \Big(\frac{{\widetilde a}_0{\widetilde\alpha}_0}{2}+\sum_{m=1}^\infty {\widetilde a}_m{\widetilde\alpha}_m\Big)+\Big(\sum_{m=1}^\infty {\widetilde b}_m{\widetilde\beta}_n\Big)=1
\end{equation}
\noindent with $\{{\widetilde\alpha}_0, {\widetilde\alpha}_n, {\widetilde\beta}_n\}$ and $\{{\widetilde a}_0, {\widetilde a}_n, {\widetilde b}_n\}$ denoting the Fourier coefficients of $f(x)$ and $g(x)$, respectively. Parseval's identity only applies to two functions. Section~\ref{PaId} presents our generalization of Parseval's identity that applies to an arbitrary number of functions.

\noindent Now, we start by solving the case of two players called {\it Alice} and {\it Bob}. To this aim, we choose $f(x)=g(x)$. Furthermore, for simplicity, we consider an even main function $f(x)$ i.e., $f(x)= f(-x)$ (so, ${\widetilde\beta}_m=0$). In this case, Parseval's identity reduces to:
\begin{equation}\label{pars4}
  \frac{{\widetilde\alpha}_0^2}{2}+\sum_{m=1}^\infty{\widetilde\alpha}_m^2=1\quad{\rm and}\quad \eta=\frac{1}{l}\int_{-l}^l \phi(x)^2dx
\end{equation}
In the sequel, in order not to burden the notations the tilde over the Fourier coefficients will be omitted being understood that the main functions are normalized. In the following Subsections, we establish the tasks of Alice and Bob.

\subsection{Tasks of Alice}

\noindent {\bf 1)} Alice splits her secret code in four parts: $x_1a=a_1+a_2+a_3+a_4$;                                                    .

\noindent {\bf 2)} Alice choses two musks, by choosing {\it arbitrary pairs of parameters}\footnote{It is convenient to choose $\omega^{(0)}_1=\alpha^{(0)}(a^{(0)}_1+ib^{(0)}_1)$ and $\omega_{1,m}=(a_1+ib_1)\alpha_m$.} : $\omega_{1,m}=a_{1,m}+ib_{1,m}$ and $\omega_1^{(0)}=a_1^{(0)}+ib_1^{(0)}$;

\noindent {\bf 3)} Let us call $(\alpha^{(0)},\alpha_m)$ the cosine Fourier components of the {\it main function} $f(x)$. We define  $(\alpha_1^{(0)},\alpha_{1,m})\equiv(|y|^{1/2}a\alpha^{(0)},|y|^{1/2} a\alpha_m)$;

\noindent {\bf 4)} Alice constructs the four hyper-vectors $A_1^{(1)}, A_1^{(2)}, B_1^{(1)}, B_1^{(2)}$, defined as
\begin{eqnarray}\label{s2D1}
  \!\!\!\!\!\!\!\!\!\!\!\!&& A_1^{(1)}\equiv\left\{a_1,\alpha_1^{(0)}+\omega_1^{(0)},\alpha_{1,m}+\omega_{1,m}\right\}\ \nonumber \\
  \!\!\!\!\!\!\!\!\!\!\!\!&& B_1^{(1)}\equiv\left\{a_3,\alpha_1^{(0)}+i\omega_1^{(0)},\alpha_{1,m}+i\omega_{1,m}\right\}\nonumber \\
  \!\!\!\!\!\!\!\!\!\!\!\!&& A_1^{(2)}\equiv\left\{a_2,\alpha_1^{(0)}-\omega_1^{(0)},\alpha_{1,m}-\omega_{1,m}\right\}\  \nonumber \\
  \!\!\!\!\!\!\!\!\!\!\!\!&& B_1^{(2)}\equiv\left\{a_4,\alpha_1^{(0)}-i\omega_1^{(0)},\alpha_{1,m}-i\omega_{1,m}\right\}\nonumber
\end{eqnarray}
\noindent {\bf 5)}  Alice sends the {\it hyper-vectors} $A_1^{(1)}$ and $A_1^{(2)}$ to the {\it node 1} and the{\it  node 2}, respectively, and the hyper-vectors $B_1^{(1)}$ and $B_1^{(2)}$ to the {\it node 3} and {\it nde 4}, respectively.

\subsection{Tasks of Bob}

\noindent {\bf 1)} Bob splits his secret code in four parts: $x_2b=b_1+b_2+b_3+b_4$;                                                    .

\noindent {\bf 2)} Bob choses two musks, by choosing {\it arbitrary pairs of parameters}\footnote{It is convenient to choose $\omega^{(0)}_2=\alpha^{(0)}(a^{(0)}_2+ib^{(0)}_2)$ and $\omega_{2,m}=(a_2+ib_2)\alpha_m$.}: $\omega_{2,m}=a_{2,m}+ib_{2,m}$ and $\omega_2^{(0)}=a_2^{(0)}+ib_2^{(0)}$;                                .

\noindent {\bf 3)} With the cosine Fourier components of the main function $f(x)$, i.e. ($\alpha^{(0)}, \alpha_m$), we define $(\alpha_2^{(0)},\alpha_{2,m})\equiv(|y|^{1/2}b\alpha^{(0)},|y|^{1/2} b\alpha_m)$;

\noindent {\bf 4)} Bob constructs the four {\it hyper-vectors} $A_2^{(1)}, A_2^{(2)}, B_2^{(1)}, B_2^{(2)}$, defined as
\begin{eqnarray}\label{s2D2}
  \!\!\!\!\!\!\!\!\!\!\!\!&& A_2^{(1)}\equiv\left\{b_1,\alpha_2^{(0)}+\omega_2^{(0)},\alpha_{2,m}+\omega_{2,m}\right\} \  \nonumber \\
  \!\!\!\!\!\!\!\!\!\!\!\!&& B_2^{(1)}\equiv\left\{b_3,\alpha^{(0)}_{2}+i\omega_2^{(0)},\alpha_{2,m}+i\omega_{2,m}\right\}  \nonumber \\
  \!\!\!\!\!\!\!\!\!\!\!\!&& A_2^{(2)}\equiv\left\{b_2,\alpha_2^{(0)}-\omega_2^{(0)},\alpha_{2,m}-\omega_{2,m}\right\} \   \nonumber \\
  \!\!\!\!\!\!\!\!\!\!\!\!&& B_2^{(2)}\equiv\left\{b_4,\alpha_2^{(0)}-i\omega_2^{(0)},\alpha_{2,m}-i\omega_{2,m}\right\}\nonumber
\end{eqnarray}
\noindent {\bf 5)}  Bob sends the hyper-vectors $A_2^{(1)}$ and $A_2^{(2)}$ to the {\it node 1} and the{\it  node 2}, respectively, and the hyper-vectors $B_2^{(1)}$ and $B_2^{(2)}$ to the {\it node 3} and {\it node 4}, respectively.

\subsection{Tasks of the Nodes}

\noindent The four nodes perform the following tasks. Note that the sign is $+$ if $y>0$ and $-$ if $y<0$:

\noindent {\it Node 1 computes}:
\begin{dmath}\label{s2d3}
  N_1= a_1+b_1\pm\Big(\frac{1}{8}(\alpha_1^{(0)}+\omega_1^{(0)})(\alpha_2^{(0)}+\omega_2^{(0)})+\frac{1}{4}\sum_{m=1}^\infty(\alpha_{1,m}+\omega_{1,m})(\alpha_{2,m}+\omega_{2,m})\Bigr)
  \nonumber
\end{dmath}
\noindent {\it Node 2 computes}
\begin{dmath}\label{s2d4}
  N_2=a_2+b_2\pm\Bigl(\frac{1}{8}((\alpha_1^{(0)}-\omega_1^{(0)})(\alpha_2^{(0)}-\omega_2^{(0)}))+\frac{1}{4}\sum_{m=1}^\infty((\alpha_{1,m}-\omega_{1,m})(\alpha_{2,m}-\omega_{2,m}))\Bigr)\nonumber
\end{dmath}

\noindent {\it Node 3 computes:}
\begin{dmath}\label{s2d5}
  N_3=a_3+b_3\pm\Bigl(\frac{1}{8}((\alpha_1^{(0)}+i\omega_1^{(0)})(\alpha_2^{(0)}+i\omega_2^{(0)}))+\frac{1}{4}\sum_{m=1}^\infty((\alpha_{1,m}+i\omega_{1,m})(\alpha_{2,m}+i\omega_{2,m}))\Bigr)\nonumber
\end{dmath}

\noindent {\it Node 4 computes}:
\begin{dmath}\label{s2d6}
  N_4=a_4+b_4\pm\Bigl(\frac{1}{8}((\alpha_1^{(0)}-i\omega_1^{(0)})(\alpha_2^{(0)}-i\omega_2^{(0)}))+\frac{1}{4}\sum_{m=1}^\infty((\alpha_{1,m}-i\omega_{1,m})(\alpha_{2,m}-i\omega_{2,m}))\Bigr)\nonumber
\end{dmath}

\subsection{Task of the Display}

\begin{figure*}[h]
  \centering
  \includegraphics[scale=.15]{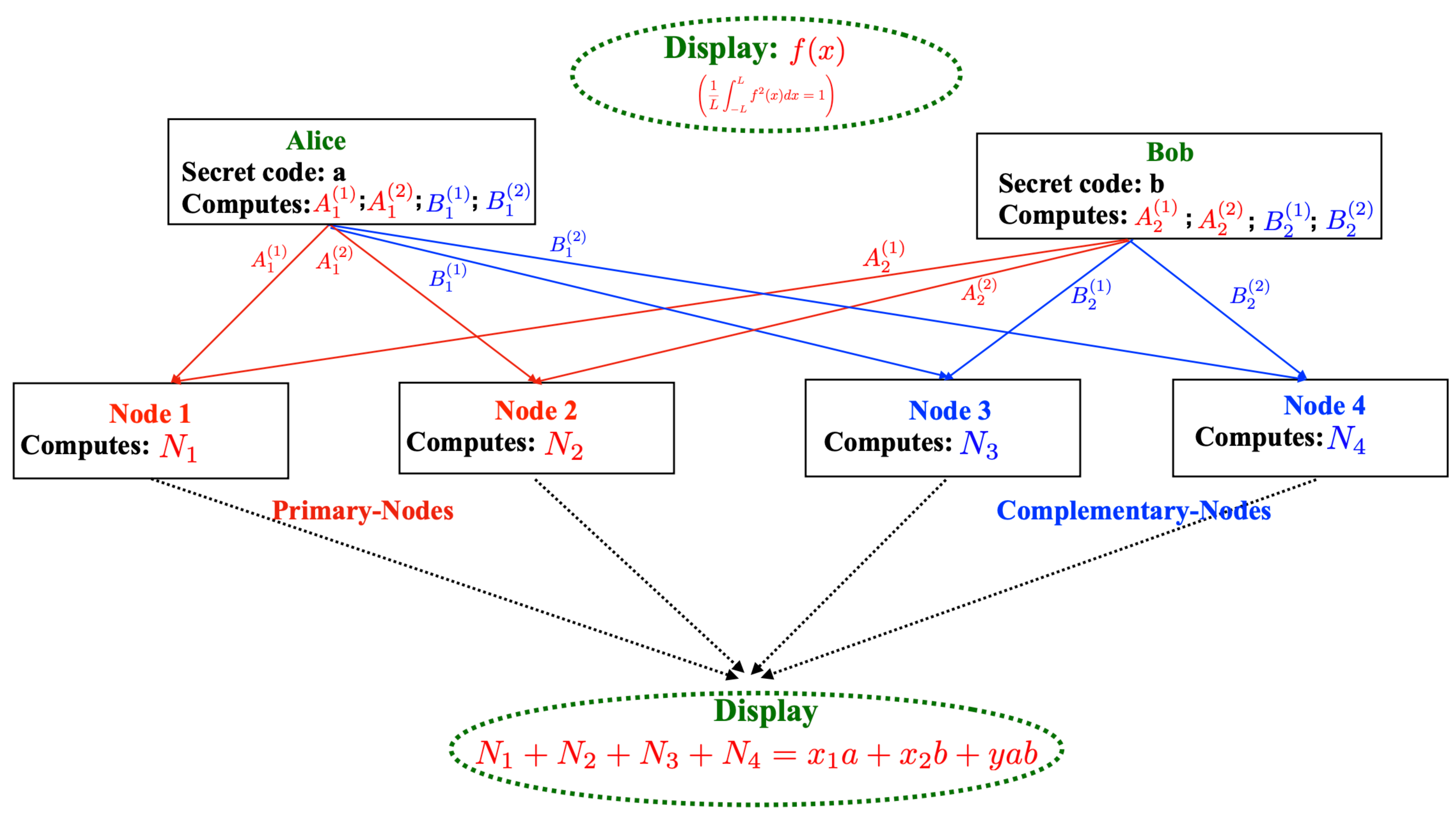}
  \caption{{\bf Display of the expression $x_1a+x_2b+yab$ by keeping secret the code $a$ and $b$.} This procedure allows showing publicly the expression by keeping secret the codes of Alice and Bob. In this case, at least three nodes must not be corrupted}
  \label{fig_crypt2D}
\end{figure*}

\noindent The display shows:
\begin{dmath}\label{s2D7}
  N_1+N_2+N_3+N_4=a_1+a_2+a_3+a_4+b_1+b_2+b_3+b_4\pm\frac{1}{2}\alpha_1^{(0)}\alpha_2^{(0)} \pm\sum_{m=1}^\infty\alpha_{1,m}\alpha_{2,m}=x_1a+x_2b+yab\nonumber
\end{dmath}
\noindent due to {\it Parseval's identity}~(\ref{pars4}). Here, at least three nodes must not be corrupted. Fig.~(\ref{fig_crypt2D}) illustrates the procedure.
\subsection{Example: $Expr=3a+5b-9ab$}
Let us apply the procedure to a concrete example where $(x_1,x_2,y)=(3,5,9)$. This triad of numbers is shown publicly. We suppose that the secret codes of Alice and Bob are $(a,b)=(2.2,4.1)$, so the value of the expression shown by the display is $Expr=-54.08$. First, we have to choose the main function $f(x)$. We choose, for example, $f(x)=\left(1+\frac{\sin(2\pi\tau)}{2\pi\tau}\right)^{-1/2}\!\!\!\!\!\!\!\!\cos(\pi\tau x/L)$. So, the Fourier coefficients read
\begin{eqnarray}\label{ex1}
  && \alpha^{(0)}=2\left(1+\frac{\sin(2\pi\tau)}{2\pi\tau}\right)^{-1/2}\frac{\sin(\pi\tau)}{\pi\tau} \\
  && \alpha^{(m)}=\alpha^{(0)}\tau^2\frac{(-1)^m}{\tau^2-m^2}\nonumber
\end{eqnarray}
\noindent If the system chooses, for example, $\tau=1/6$, we get
\begin{eqnarray}\label{ex2}
  && \alpha^{(0)}=6\sqrt{\frac{2}{3\sqrt{3}\pi+2\pi^2}} \\
  && \alpha_m=6\sqrt{\frac{2}{3\sqrt{3}\pi+2\pi^2}}\frac{(-1)^m}{1-36m^2}\nonumber
\end{eqnarray}
\noindent It is useful to take into account the following identity \cite{gradshteyn2014table}\footnote{Useful relations for getting the values of the infinite sums can be found in Appendix~\ref{A}.}
\begin{equation}\label{s2D10}
  \sum_{m=1}^\infty(\alpha_m)^2=\frac{-36+3\sqrt{3}\pi+2\pi^2}{\pi(3\sqrt{3}+2\pi)}\nonumber
\end{equation}
\subsubsection{Tasks of Alice}
\noindent Alice possesses the code $a=2.2$. She splits her ode in four pieces $a_1=3.3\ ;\ a_2=1.65\; \ a_3=1.32\ ;\ a_4=0.33$ and chooses the following masks
\begin{equation}\label{ex3}
  \omega_1^{(0)}=\alpha^{(0)}(7+9i)\quad ; \quad \omega_1=(2+11i)\alpha_m\nonumber
\end{equation}
\subsubsection{Tasks of Bob}
\noindent Bob possesses the code $a=4.1$. He splits his code in four pieces $b_1=3.41667\ ;\ b_2=2.05\; \ b_3=5.125\ ;\ b_4=9.90833$ and chooses the following masks
\begin{equation}\label{ex4}
  \omega_2^{(0)}=\alpha^{(0)}(5+3i)\quad ; \quad \omega_2=(4+8i)\alpha_m\nonumber
\end{equation}
\subsubsection{Tasks of the Nodes}
\noindent The nodes perform the following tasks:

\noindent {\it Node 1}:
\begin{dmath}\label{ex5}
  N_1\!=\! a_1\!+\!b_1 - \frac{1}{8}(\alpha_1^{(0)}\!+\!\omega_1^{(0)})(\alpha_2^{(0)}\!+\!\omega_2^{(0)})\!+\!\frac{1}{4}\sum_{m=1}^\infty(\alpha_{1,m}\!+\!\omega_{1,m})(\alpha_{2,m}+\omega_{2,m}) =\!3.3\!+\!3.41667\!-\!\! 1/8(\alpha^{(0)})^2(13.6\!+\!9i)(17.3\!+\!3i)\!-\!1/4(8.6\!+\!11i)(16.3\!+\!8i)\!\!\sum_{m=1}^\infty\!\!(\alpha_m)^2
  \nonumber
\end{dmath}
\noindent {\it Node 2}:
\begin{dmath}\label{ex6}
  N_2= a_2\!+\!b_2\!-\!\left(\frac{1}{8}((\alpha_1^{(0)}\!-\!\omega_1^{(0)})(\alpha_2^{(0)}-\omega_2^{(0)}))\!+\!\frac{1}{4}\!\! \sum_{m=1}^\infty\! ((\alpha_{1,m}\!-\!\omega_{1,m})(\alpha_{2,m}\!-\!\omega_{2,m}))\right)  = 11.1887 + 16.153 i\nonumber
\end{dmath}
\noindent  {\it Node 3}:
\begin{dmath}\label{ex7}
  N_3\!=\!a_3\!+\!b_3\!-\!\left(\frac{1}{8}((\alpha_1^{(0)}\!+\!i\omega_1^{(0)})(\alpha_2^{(0)}\!+\!i\omega_2^{(0)}))\!+\!\frac{1}{4}\!\! \sum_{m=1}^\infty \! ((\alpha_{1,m}\!+\!i\omega_{1,m})(\alpha_{2,m}\!+\!i\omega_{2,m}))\!\right) =20.7616 - 13.2477 i \nonumber
\end{dmath}
\noindent {\it Node 4}:
\begin{dmath}\label{ex8}
  N_4\!=\!a_4\!+\!b_4\!-\!\left(\frac{1}{8}((\alpha_1^{(0)}\!-\!i\omega_1^{(0)})(\alpha_2^{(0)}\!-\!i\omega_2^{(0)}))\!+\!\frac{1}{4}\!\! \sum_{m=1}^\infty((\alpha_{1,m}\!-\!i\omega_{1,m})(\alpha_{2,m}\!-\!i\omega_{2,m}))\!\right)
  =-40.7457 + 46.2424 i \nonumber
\end{dmath}
\subsubsection{Tasks of the Display}
\noindent The display shows publicly:
\begin{equation}\label{ex9}
  Expr=N_1+N_2+N_3+N_4=-54.08 + 7.10543\cdot 10^{-15} i\nonumber
\end{equation}
\noindent due to {\it Parseval's identity}~(\ref{pars4}).

\section{Showing a general mathematical expression by keeping secret the codes $a$ and $b$}\label{GE}
It is easy to convince ourselves that the method of the masks may successfully be applied to get a wide variety of mathematical expressions, by keeping secret the code of the players and without having to resort to any mathematical approximation. For example, it is possible to get the value of
\begin{equation}\label{GE1}
  {\rm Expr.}=\arctan\log\sin(a^2+3ab-4/b)\nonumber
\end{equation}
\noindent without having to approximate it with polynomial interpolations (or another kind of interpolations). The strategy is {\it to mask} the terms and ask the Display to show the value of the mathematical expression. In the above case, 3 terms must be masked through the use of 3 masks, one provided by player $a$ and two provided by player $b$, for getting the expressions:
\begin{equation}\label{GE2}
  arg.= a^2+3ab-4/bi\nonumber
\end{equation}
\noindent and request that the display publicly shows the value of the final expression:
\begin{equation}\label{GE3}
  {\rm Expr.}=\arctan\log\sin (arg.)\nonumber
\end{equation}
\noindent Of course, the above considerations apply to any mathematical expression, of the type:
\begin{equation}\label{GE4}
  {\rm Expr.}=f(arg.)\nonumber
\end{equation}
\noindent where $f$ is a publicly visible function. However, in several practical applications, it is not possible to dispose of the exact mathematical expression of a variable, but only a discrete set of experimental sampling of it. In these cases, the method of the mask functions can still be successfully used. The {\it musk-technique} is able to evaluate the following expression, by keeping secret the codes $a$ and $b$:
\begin{equation}\label{GE5}
  {\rm Expr.}= w_1(a)+w_2(b)+\sum_{r,p=0}^{m,m'}c_{{r}p}g_{r}(a)h_p(b)
\end{equation}
\noindent where $w_1$, $w_2$, $g_r$ and $h_p$ are functions and $m$ and $m'$ (finite) integers, respectively. $c_{lp}$ are constant coefficients. In this case, both Alice and Bob have to split functions $w_1$, and $w_2$ into two parts and they have to construct $m$ and $m’$ mask functions, respectively. As known, any regular function defined between $-1$ and $+1$ can be approximated, with high precision, considering only a few terms of the {\it Chebyshev polynomials}. This observation allows to approximate any cryptographic expression with the first (no more than, let's say the first nine) polynomials of {\it Chebyshev} and, then, to apply the musk-functions technique to the expression obtained with the {\it Chebyshev polynomials}. Note that the accurate convergence of the {\it Chebyshev polynomials} is guaranteed in the interval of the type $-l$ and $+l$ (if the polynomials are orthonormalized). $l$ is an arbitrary parameter that does not participate in the encryption procedure. So, it may be chooses $l = 1$. Thanks to {\it Chebyshev polynomials}, the general expression may be brought into the form (\ref{GE5}). This will be more precise in the forthcoming sections.

\subsection{Properties and Theorems on the Chebyshev Polynomials }
It is useful to recall the two main {\it properties} and {\it theorems} on the Chebyshev Polynomials\footnote{A deep and exhaustive analysis on the Chebyshev polynomials can be found, for instance, in \cite{mason}.}.

\noindent A) {\it Properties of the Chebyshev polynomials}

\noindent i) The Chebyshev polynomials $T_m(x)$ form a complete orthogonal system.

\noindent ii) The Chebyshev series converges to $\Psi(x)$ if the function is piecewise smooth and continuous. The smoothness requirement
can be relaxed in most cases – as long as there are a finite number of discontinuities in $\Psi(x)$ and its derivatives.

\noindent iii) At a discontinuity, the series will converge to the average of the right and left limits.

\noindent B) {\it Theorems on the Chebyshev polynomials}

\noindent {\bf Theorem 1: Accuracy}

\noindent {\it If we want polynomial interpolating a function $f$ at $m+1$ points $x_s$ in the interval $[-1,1]$ to be as accurate as possible, then we
  should choose the data $x_s$ so that they are the zeros of the Chebyshev polynomial} $T_{m+1}(x)$. More specifically, {\it If the nodes  $x_s$ are chosen as the roots of the Chebyshev polynomial} $T_{m+1}(x)$
\begin{equation}\label{C1}
  x_s=\cos\left(\frac{2s+1}{2m+2}\pi\right)\quad ;\quad (s=0,1\dots,m)\nonumber
\end{equation}
\noindent {\it then the error term for polynomial interpolation using the nodes $x_s$ is}
\begin{equation}\label{C2}
  E(x)=|f(x)-P(x)|\leq\frac{1}{2^m(m+1)!}{\rm max}_{(-1\leq t\leq 1)}|f^{m+1}(t)|\nonumber
\end{equation}
\noindent {\it Moreover, this is the best upper bound we can achieve by varying the choice of} $x_s$.

\noindent {\bf Theorem 2: Convergence}

\noindent {\it The Chebyshev Numerical Method is best with the Rate of Convergence}.

\noindent The different numerical methods have different rates of convergence. And the rate of convergence is a very important issue in the solution of polynomial and transcendental equations because the rate of convergence of any numerical method determines the speed of the approximation to the solution of the problem. The following table shows the comparison of the rate of convergence. Among Secant, {\it Regula-Falsi }and {\it Newton-Raphson} which are based on 1st-degree equations Newton-Raphson has a good rate of convergence i.e. $2$. Among {\it Muller} and {\it Chebyshev methods}, which are based on $2^{nd}$ degree equations, Chebyshev is best with the rate of convergence of $3$. Table~\ref{tab_Cheb_Conv} compares the rate of convergence of various methods.

\begin{table}[t]
  \centering
  \footnotesize
  \begin{tabular}{lrrrr}
    \toprule
    Method         & Based on Equation & Rate of Convergence \\
    \midrule
    Secant         & $1^{st}$ degree   & 1.618               \\
    Regula-Falsi   & $1^{st}$ degree   & 1                   \\
    Newton-Raphson & $1^{st}$ degree   & 2                   \\
    Muller         & $2^{st}$ degree   & 1.84                \\
    Chebyshev      & $2^{st}$ degree   & 3                   \\
    \bottomrule
  \end{tabular}
  \caption{\label{tab:widgets}Rate of convergence of various methods}
  \label{tab_Cheb_Conv}
\end{table}

\subsection{Interpolation of a General Expression $Expr. (a,b)$ with the Chebyshev Polynomials}

Let $Expr.(a,b)=w_1(a)+w_2(b)+\Phi(a,b)$ be a general function of two variables $a$ and $b$. We may interpolate this expression by proceeding in the following manner:

\noindent {\bf i)} We develop the expression $\Phi(a,b)$ in terms of Chebyshev polynomials $T_{r}(x)$ with respect to the variable $a$:
\begin{equation}\label{C3}
  \Phi(a,b)\simeq\sum_{r=0}^m T_r(a)c_{r}(b)\nonumber
\end{equation}
\noindent {\bf ii)} Successively, we develop the expressions $c_{{r}}(b)$ in terms of Chebyshev polynomials $T_p(x)$ with respect to the variable $b$:
\begin{equation}\label{C4}
  c_{r}(b)\simeq\sum_{p=0}^{m'}c_{rp}T_p(b)\nonumber
\end{equation}
\noindent Finally, we get
\begin{equation}\label{C5}
  {\rm Expr.}\simeq w_1(a)+w_2(b)+\sum_{r,p=0}^{m,m'}c_{rp}T_r(a)T_p(b)
\end{equation}
\noindent which is of the same form as Eq.~(\ref{GE5}). To sum up, the main conclusions of our analysis are:

\noindent {\bf a)} Theorems {\bf 1)} and {\bf 2)} ensure the {\it best accuracy} - by quantifying the error - and the {\it best rate of convergence};

\noindent {\bf b)} The previous development can trivially be extended when {\it the expression depends on n variables}.

\section {Mathematical Framework for Treating the Case of Multiple Users}\label{math}
In the previous Sections, we solved the problem for the case of two users. Unfortunately, the adopted procedure does not trivially extend to the case of $n$ users (with $n> 2$). This is due to the following two drawbacks:

\noindent {\bf a)} Parseval's identity traditionally applies only to two functions. In this case, the mathematical expression of this identity is invariant under the permutation of the two functions; in other words, the two functions are {\it indistinguishable} each with the other. However, the problem arises when the users are more than two: if we apply this identity in a simple sequential manner, the functions will no longer be indistinguishable from each other. What we need is to derive an expression equivalent to Eq.~(\ref{pars1}), which applies for $n\geq 2$ functions. In this way, the property of indistinguishability among functions will be preserved. Concretely, we need to derive the expression for the normalisation coefficient $\eta$ that generalises Eq.~(\ref{pars2}) for $n\geq 2$ functions, so that Parseval's identity for normalised functions can be cast into the form (\ref{pars3}). This task will be accomplished in the next Section.

\noindent {\bf b)} The second obstacle is due to the fact that the algebra of complex numbers, which we used to treat the case $n = 2$, turns out to be inadequate when there are more than two users. Indeed, for $n>2$ the additive masks no longer cancel each other leading to an erroneous final result. We may easily convince ourselves on this by noticing that for $n = 2$ the masks cancel each with other  because $i^2 = -1$,  while for $n>2$ ($n = 4$, for example) we have $i^n = + 1$ for $n=4\iota$, with $\iota$ denoting a natural number. In this latter case, the masks instead of canceling each other add up! This issue will be overcome by introducing an appropriate algebra, referred to as {\it Theta algebra}, which has the property to allow the cancellation of the masks for $n\geq 2$, thus providing the correct final result (see the forthcoming Subsection).

\subsection{Generalisation of Parseval’s identity for $n$ functions}\label{PaId}

\begin{figure}[htb]
  \centering
  \includegraphics[width=5cm,height=8cm]{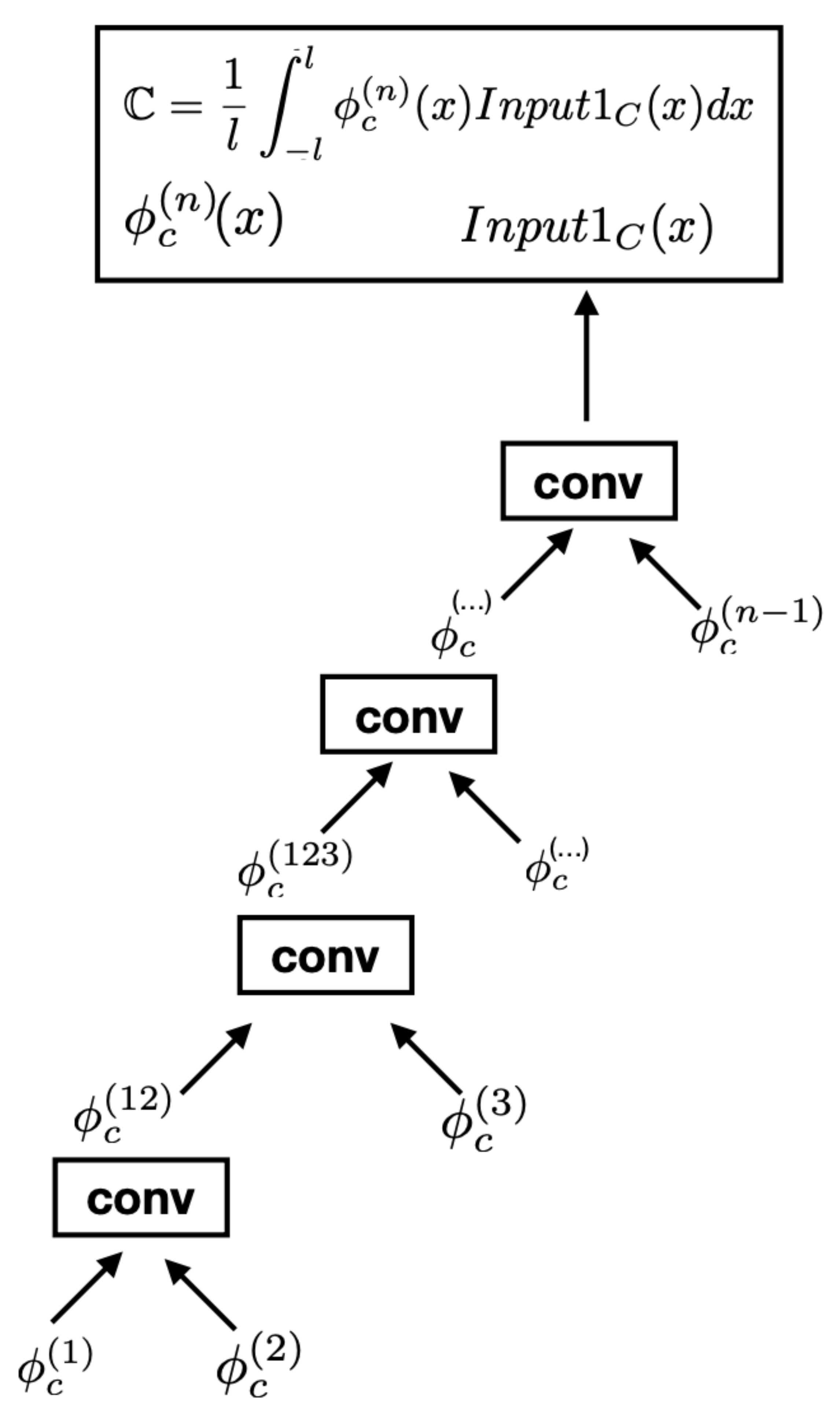}
  \caption{
    Illustration of the algorithm for calculating constant $\mathbb C$. $n-2$ convolution operations are performed sequentially starting from $\phi_c^{(1)}(x)$ to $\phi_c^{(n-1)}(x)$ i.e. until we get only one function. This latter function has to be integrated with $\phi_c^{(n)}(x)$. Note that, due to the commutative property, the order of the input functions is not relevant in the convolution operations.
    \label{fig_C}}
\end{figure}

\begin{figure}[htb]
  \centering
  \includegraphics[width=8cm,height=8cm]{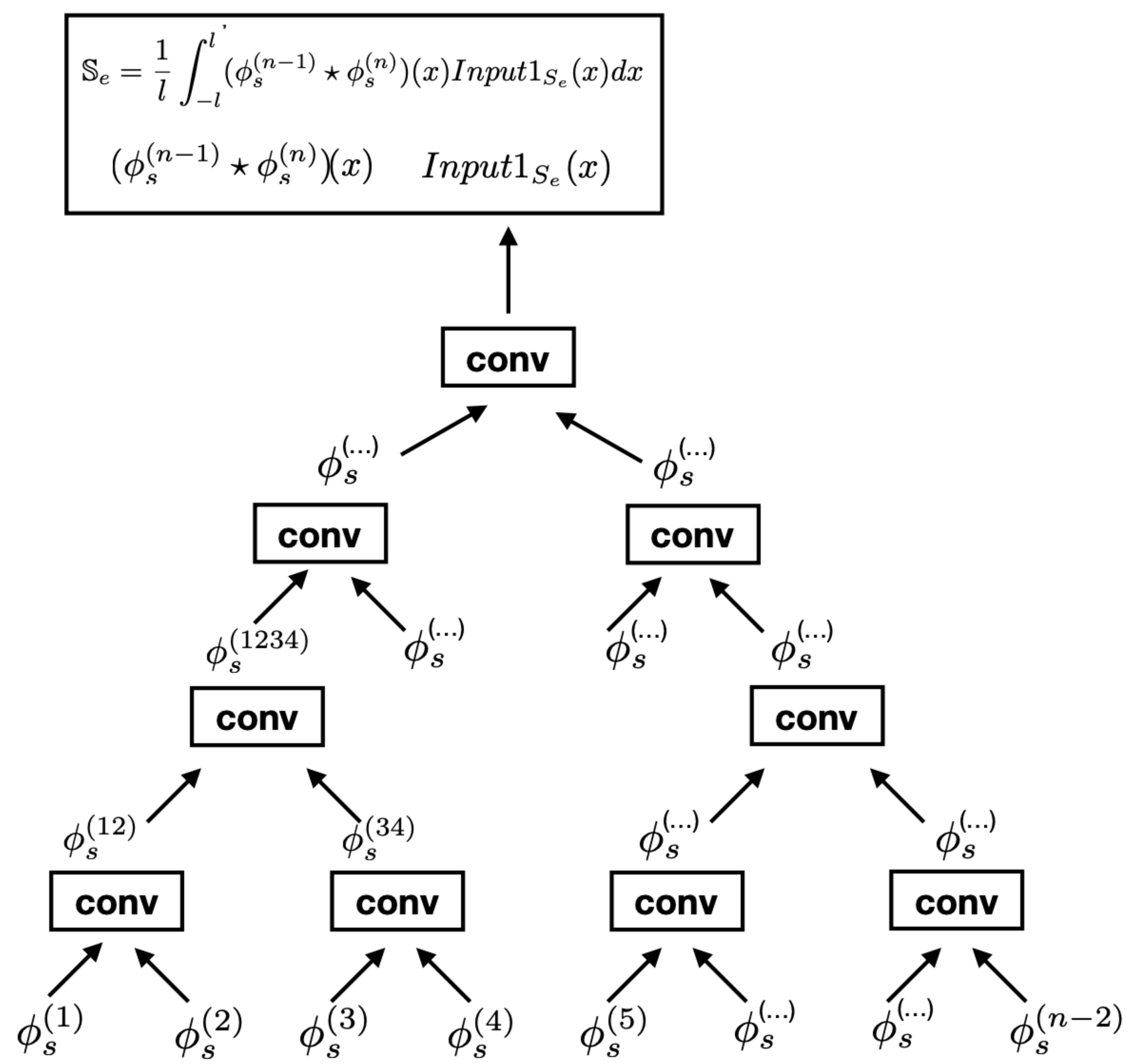}
  \caption{
    Illustration of the algorithm for calculating constant ${\mathbb S}_e$. If $n$ is an even number, we pair up two by two the $n-2$ input functions and we perform $n-3$ convolution operations between each pair. We iterate this process until there remains only one final function. This latter function is finally integrated with the convolution operation between the function $\phi_s^{(n-1)}$ and $\phi_s^{(n)}$ [i.e., $(\phi_s^{(n-1)}\star\phi_s^{(n)})(x)$]. Due to the commutative property, the order of the input functions is not relevant in the convolution operations.\label{fig_Se}}
\end{figure}

\begin{figure}[htb]
  \centering
  \includegraphics[width=8cm,height=8cm]{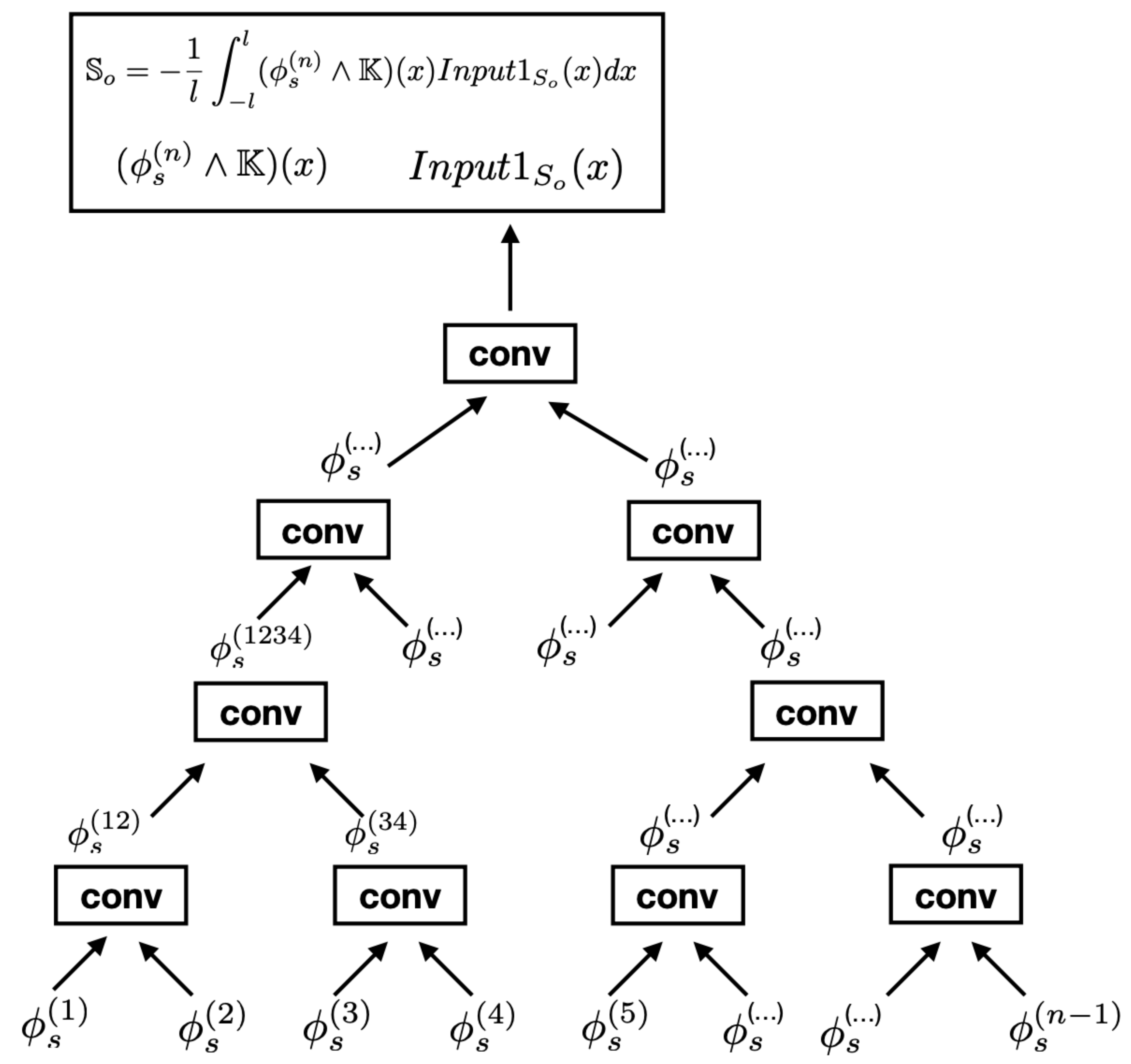}
  \caption{
    Illustration of the algorithm for calculating constant ${\mathbb S}_o$. If $n$ is an odd number, we pair up two by two the $n-1$ input functions and we perform $n-3$ convolution operations between each pair. We iterate this process until there remains only one final function is finally integrated. This latter function has to be integrated with function $(\phi_s^{(n)}\wedge{\mathbb K}(x))(x)$. Due to the commutative property, the order of the input functions is not relevant in the convolution operations.
    \label{fig_So}}
\end{figure}

In this section, we present the generalization of Parseval's identity for the Fourier series applicable to $n$ inputs. Here, we will limit ourselves only to enunciating the theorem, delegating the proof of the theorem in Appendix~\ref{B}. With the $n$ main functions $\phi^{(j)}(x)$, with ($j=1,\cdots , n$), we construct out $n$ {\it even functions} and $n$ {\it odd functions}:
\begin{eqnarray}\label{pai1}
  && \phi_c^{(j)}(x)\equiv\frac{1}{2}\Bigl(\phi^{(j)}(x)+\phi^{(j)}(-x)\Bigr)\ \nonumber \\
  && \phi_s^{(j)}(x)\equiv\frac{1}{2}\Bigl(\phi^{(j)}(x)-\phi^{(j)}(-x)\Bigr)
\end{eqnarray}
\noindent with $ (j=1,2,\cdots)$. We also introduce the {\it convolution operation} ($\star$) between two functions $\phi(x)$ and $\psi(x)$:
\begin{equation}\label{pai2}
  \bigl(\phi\star\psi\bigr)(x)=\frac{1}{l}\int_{-l}^{l}\phi(t)\psi(x-t)dt\nonumber
\end{equation}
\noindent and the {\it integral transform} ($\wedge$) of a function $\phi(x)$ by a {\it kernel function} of two variable ${\mathbb K}(x,t)$, defined as:
\begin{eqnarray}\label{pai3}
  &&\bigl(\phi\wedge{\mathbb K}\bigr)(x)=\frac{1}{l}\int_{-l}^l\phi(t){\mathbb K}(x,t)dt\quad{\rm with} \nonumber \\
  &&{\mathbb K}(x,t)=\sum_{m=1}^\infty \cos\Bigl(\frac{m\pi}{l}x\Bigr)\sin\Bigl(\frac{m\pi}{l}t\Bigr) \nonumber \\
  && =\frac{\sin(\pi t/l)}{2(\cos(\pi x/l)-\cos(\pi t/l))}\nonumber
\end{eqnarray}
\noindent Note that the Kernel ${\mathbb K}(x,t)$ should be intended in a {\it distribution} sense \cite{infPS}. Let us now consider $n$ inputs (main functions) with the Fourier representations
\begin{dmath}\label{pai4}
  \phi^{(j)}(x)=\frac{\alpha_0^{(j)}}{2}+\sum_{m=1}^{+\infty}\alpha_m^{(j)}\cos\Bigl(\frac{m\pi x}{l}\Bigr)+\sum_{m=1}^{+\infty}\beta_m^{(j)}\sin\Bigl(\frac{m\pi x}{l}\Bigr)\ \ {\rm with}\ \ j=1,2,\cdots, n \nonumber
\end{dmath}
\noindent We introduce the following three constants
\begin{eqnarray}\label{pai5}
  &&{\mathbb C}= \frac{1}{l}\int_{-l}^l\phi_c^{(n)}(x) Input1_C(x)dx\\
  &&{\mathbb S}_e=\frac{1}{l}\int_{-l}^l(\phi_s^{(n-1)}\star\phi_s^{(n)})(x) Input1_{S_e}(x)dx\nonumber\\
  &&{\mathbb S}_o=-\frac{1}{l}\int_{-l}^l(\phi_s^{(n)}\wedge{\mathbb K})(x)Input1_{S_o}(x)dx\nonumber
\end{eqnarray}
\noindent where functions $Input1(x)$ are defined as
\begin{dmath}\label{pai5a}
  Input1_C(x)\equiv((\cdots (((\phi_c^{(1)}\star\phi_c^{(2)})\star\phi_c^{(3)})\star\phi_c^{(4)})\star\cdots)\star\phi_c^{(n-1)}\bigr)(x)
\end{dmath}
\begin{dmath}
  Input1_{S_e}(x)\equiv\bigl((\phi_s^{(1)}\star\phi_s^{(2)})\star(\phi_s^{(3)}\star\phi_s^{(4)})\star\cdots\star(\phi_s^{(n-3)}\star\phi_s^{(n-2)})\bigr)(x)\end{dmath}
\begin{dmath}
  Input1_{S_o}(x)\equiv\bigl((\phi_s^{(1)}\star\phi_s^{(2)})\star(\phi_s^{(3)}\star\phi_s^{(4)})\star\cdots\star(\phi_s^{(n-2)}\star\phi_s^{(n-1)})\bigr)(x)\nonumber
\end{dmath}
\noindent Eq.~(\ref{pai5}) may conveniently be cast into the form (see Appendix~(\ref{B})):
\begin{eqnarray}\label{pai6}
  &&{\mathbb C}=\frac{1}{2}\Pi_{j=1}^{n}\alpha_0^{(j)}+\sum_{m=1}^{\infty}\bigl(\Pi_{j=1}^{n}\alpha_m^{(j)}\bigr)\nonumber\\
  &&{\mathbb S}_\kappa=\sum_{m=1}^{\infty}\bigl(\Pi_{j=1}^n\beta_m^{(j)}\bigr)\quad {\rm with} \nonumber \\
  && \qquad{\mathbb S}_\kappa=
  \left\{ \begin{array}{ll}
    {\mathbb S}_e \  & \ \mbox{{\rm if}\ \ n={\rm even\ number}} \\
    {\mathbb S}_o\   & \ \mbox{{\rm if}\ \ n={\rm odd\ number}}  \\
  \end{array}
  \right.\nonumber
\end{eqnarray}
\noindent Hence, the generalised Parseval's identity for $n$ inputs $\phi^{(j)}(x)$ (with $j=1,2,\cdots,n$) reads:
\begin{eqnarray}\label{app7}
  &&{\mathbb C}+{\mathbb S}_\kappa \\
  && =\frac{1}{2}\Pi_{j=1}^n\alpha_0^{(j)}+\sum_{m=1}^\infty\bigl(\Pi_{j=1}^n\alpha_m^{(j)}\bigr)+\sum_{m=1}^\infty\bigl(\Pi_{j=1}^n\beta_m^{(j)}\bigr)  \nonumber \\
  &&\quad {\rm with} \; {\mathbb S}_\kappa=
  \left\{ \begin{array}{ll}
    {\mathbb S}_e \  & \ \mbox{{\rm if}\ \ n={\rm even\ number}} \\
    {\mathbb S}_o\   & \ \mbox{{\rm if}\ \ n={\rm odd\ number}}  \\
  \end{array}
  \right.
  \nonumber
\end{eqnarray}
\noindent Hence, the {\it normalisation constant} $\eta$ is given by
\begin{equation}\label{pai8}
  \eta=({\mathbb C}+{\mathbb S}_\kappa)^{-1/n}
  \ {\rm with}\ \ {\mathbb S}_\kappa=
  \left\{ \begin{array}{ll}
    {\mathbb S}_e & \mbox{{\rm if}\ n={\rm even\ number}} \\
    {\mathbb S}_o & \mbox{{\rm if}\ n={\rm odd\ number}}  \\
  \end{array}
  \right.
\end{equation}
\noindent and the {\it normalised main functions} $f^{(j)}(x)$ read
\begin{equation}\label{pai9}
  f^{(j)}(x)=\eta\phi^{(j)}(x)\quad{\rm with}\quad j=1,2,\cdots, n
\end{equation}
\noindent If we use the main functions $f^{(j)}(x)$ the generalised Parseval's identity reads
\begin{equation}\label{pai10}
  \frac{1}{2}\Pi_{j=1}^n{\widetilde\alpha}_0^{(j)}+\sum_{m=1}^\infty\bigl(\Pi_{j=1}^n{\widetilde\alpha}_m^{(j)}\bigr)+\sum_{m=1}^\infty\bigl(\Pi_{j=1}^n{\widetilde\beta}_m^{(j)}\bigr)=1
\end{equation}
\noindent with ${\widetilde\alpha}_0^{(j)}$, ${\widetilde\alpha}_m^{(j)}$, and ${\widetilde\beta}_m^{(j)}$ denoting the Fourier coefficients of $f^{(j)}(x)$. As we can see, users in the generalized Parseval's identity are indistinguishable from each other. The schemes of calculation of the coefficients ${\mathbb C}$, ${\mathbb S}_e$, and ${\mathbb S}_o$ are illustrated in Figure \ref{fig_C}, \ref{fig_Se} and \ref{fig_So}, respectively.


\subsection{The $\Theta^{[n]}$-Algebra}\label{TA}

\begin{figure*}[htb]
  \centering
  \includegraphics[width=8cm,height=5cm]{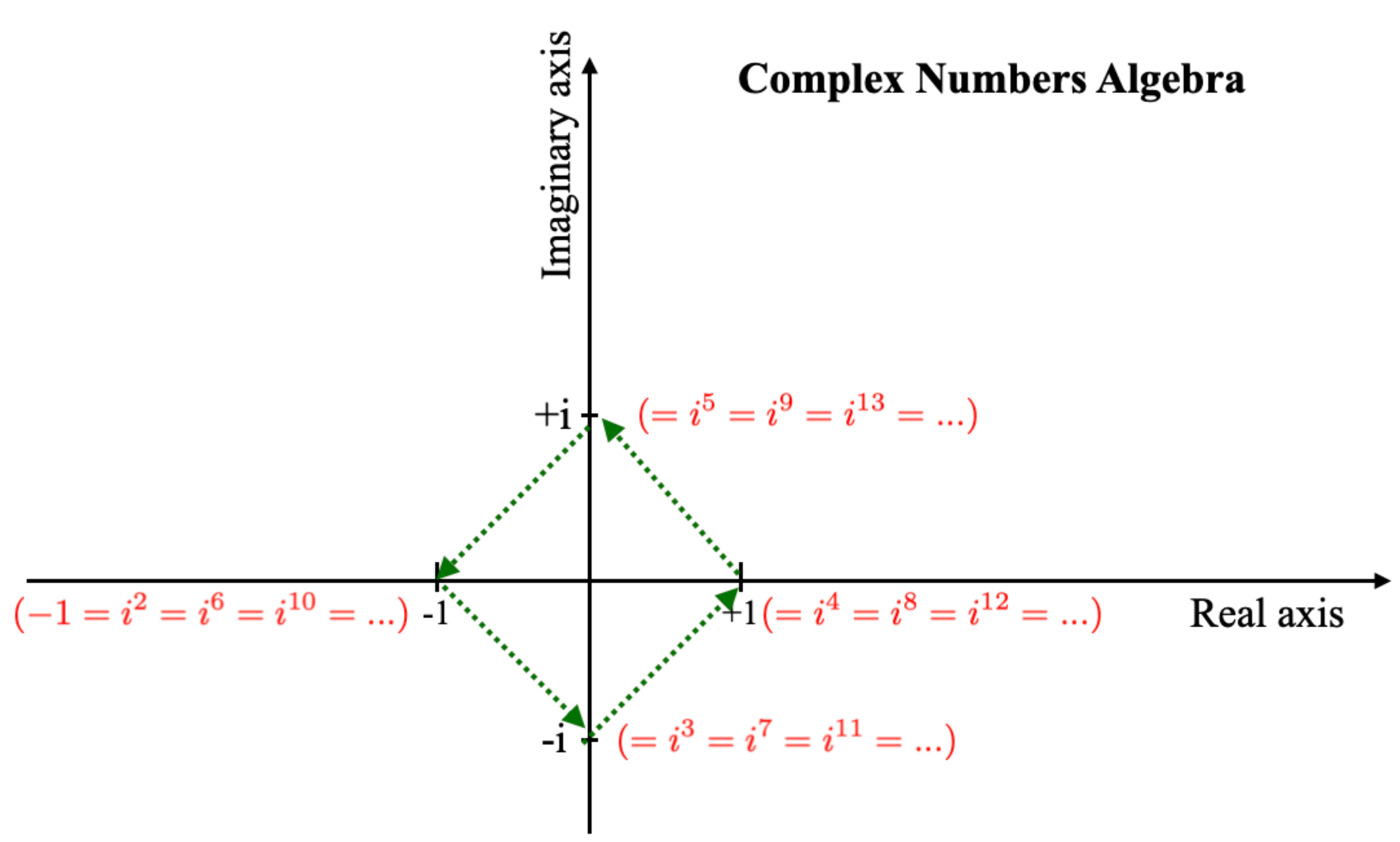}
  \caption{
    The path is followed in the complex plane by raising to power the imaginary number $i$.
    \label{fig_TA1}}
\end{figure*}

We have already noted that the algebra of complex numbers is not adequate to solve the problem when the number of users is greater than two.  Indeed, it is easily checked that in case of multiple users we are unable to make the masks disappear, and this is because, according to the algebra of complex numbers, we have
\begin{equation}\label{TA1}
  +1=i^4=i^8=i^{12}=...\quad {\rm and}\quad -i=i^3=i^7=i^{11}=...\nonumber
\end{equation}
\noindent So, by raising the imaginary number $i$ to powers of integers, we follow the path illustrated in Figure~\ref {fig_TA1}.
\noindent Our masks must then be constructed by using an algebra where the imaginary number $i$ is replaced by another number, let's call it $\zeta^{(1)}$, satisfying the identities
\begin{eqnarray}\label{TA2}
  -1=\zeta^{(2)}=\zeta^{(4)}=\zeta^{(6)}=...\quad \nonumber\\
  {\rm and}\quad+i=\zeta^{(1)}=\zeta^{(3)}=\zeta^{(5)}=\zeta^{(7)}=...\nonumber
\end{eqnarray}
\noindent In other terms, we would need to follow the path shown in Figure~\ref {TA2}.
\begin{figure*}[htb]
  \centering
  \includegraphics[width=8cm,height=5cm]{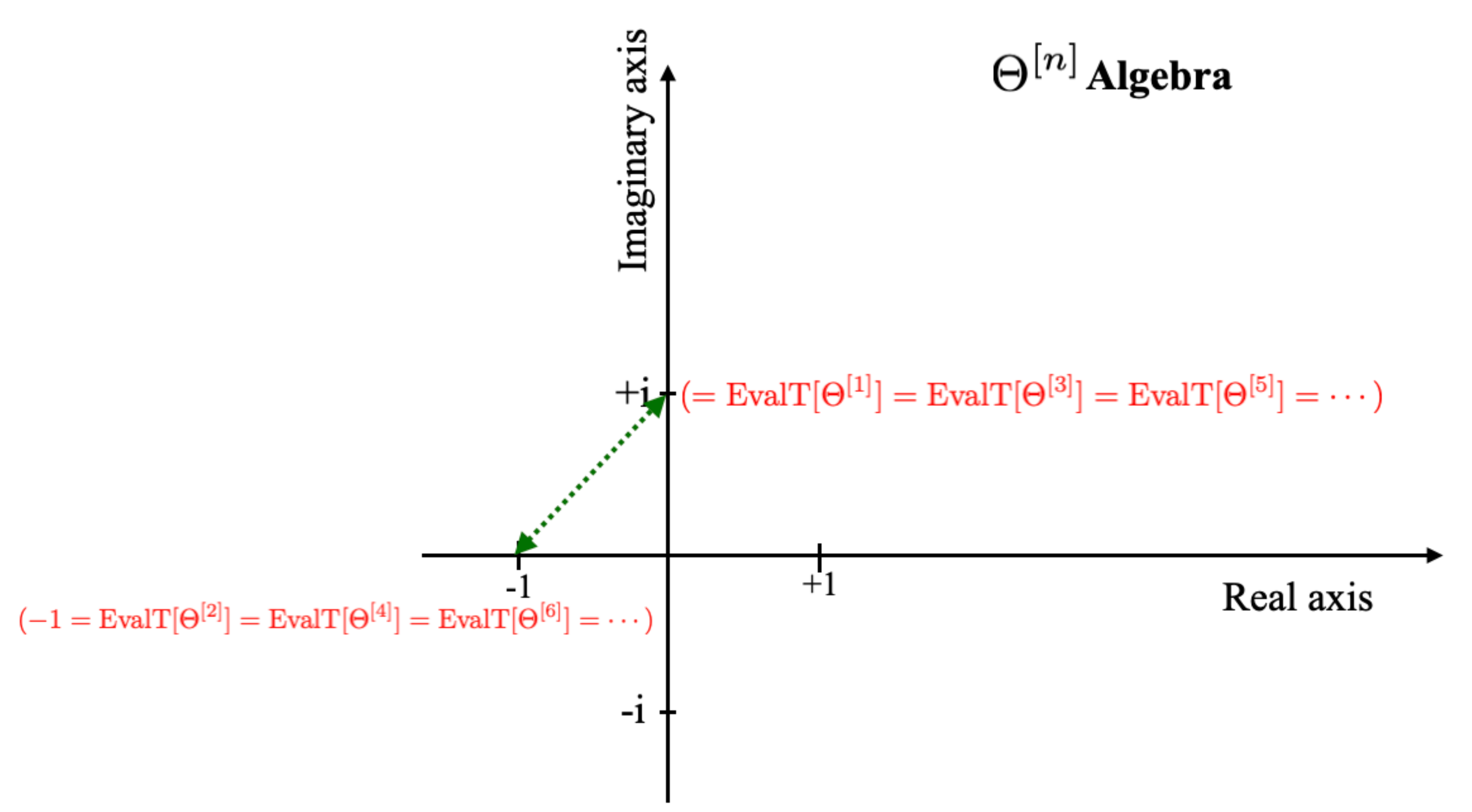}
  \caption{
    Path in the complex plane obtained by raising $\zeta^{(1)}$ to a power.
    \label{fig_TA2}}
\end{figure*}

\noindent This algebra is referred to as the $\Theta^{[n]}$-algebra and $\zeta^{(n)}\equiv{\rm EvalT}[\Theta^{[n]}]$ with ${\rm EvalT}[\Theta^{[n]}]$ denoting the {\it numerical evaluation of} $\Theta^{[n]}$. All of this will be specified in the following Subsection.

\subsection{Rules of the $\Theta^{[n]}$-algebra}
\noindent In the previous section we introduced $\Theta^{[n]}$ and the symbol ${\rm EvalT}[\Theta^{[n]}]$. More precisely,

\noindent {\bf a)} The {\it Theta-algebra} is an algebraic structure that deals with elements denoted by $\Theta^{[n]}$ and is equipped by an operation denoted by $\ast$;

\noindent {\bf b)} $\Theta^{[n]}$ is an abstract symbol satisfying a finite set of axioms reported in Appendix~\ref{C};

\noindent {\bf c)} ${\rm EvalT}[\Theta^{[n]}]$ is a command that assigns a complex number to mathematical expressions involving $\Theta^{[n]}$ according to the definition shown in Appendix~\ref{C}.

\noindent For easy reference, and in order not to burden the reading of the work, here we report only the main algorithms for $\Theta^{[n]}$ and ${\rm EvalT}[\Theta^{[n]}]$ In fact, these are the main rules we need to perform our calculations. In Appendix \ref{C} we can find a more complete set of rules to be satisfied by $\Theta^{[n]}$ and ${\rm EvalT}[\Theta^{[n]}]$ with some examples of calculation.

\begin{eqnarray}\label{TA3}
  \!\!\!\!\!\!\!\!&& \Theta^{[1]}\ast\Theta^{[1]}\ast\Theta^{[1]}\ast\cdots\ast\Theta^{[1]} \equiv\Theta^{[n]}\quad (n\ {\rm times})\nonumber \\
  \!\!\!\!\!\!\!\!&& \Theta^{[m]}\ast\Theta^{[n]}=\Theta^{[n]}\ast\Theta^{[m]}=\Theta^{[m+n]}
\end{eqnarray}
\begin{dmath}
  {\rm EvalT}[\Theta^{[n]}]\equiv\cos\left(\frac{\pi}{4}\left(3+(-1)^n\right)\right)+i\sin\left(\frac{\pi}{4}\left(3+(-1)^n\right)\right)
\end{dmath}
\begin{dmath}
  ({\rm EvalT}[\Theta^{[n]}])^m\equiv\cos\left(\frac{m\pi}{4}\left(3+(-1)^n\right)\right)+i\sin\left(\frac{m\pi}{4}\left(3+(-1)^n\right)\right)\nonumber
\end{dmath}
\noindent Hence, as desired
\begin{eqnarray}\label{TA4}
  &&{\rm EvalT}[\Theta^{[n]}]=-1\quad {\rm if}\ n\ {\rm is\ an\ even\ number}\nonumber\\
  &&{\rm EvalT}[\Theta^{[n]}]=+i \quad {\rm if}\ n\ {\rm is\ an\ odd\ number}\nonumber
\end{eqnarray}

\section{Display of the expression $\sum_{j=1}^nx_ja_j+y\Pi_{j=1}^na_j$ by keeping secret the codes $a_j$}\label{nU}
We are now in a position to solve our problem for $n$ users. First of all, without compromising the participants' privacy, we choose the same function for all users (i.e., $\phi^{(j)}(x)=\phi(x)$ $\forall j$) where $\phi(x)$ is an even function i.e., $\phi(x)=\phi(-x)$. In this case, we have ${\mathbb S}_e={\mathbb S}_o=0$ and the generalized Parseval's identity reads\footnote{In order not to burden the notations, the tilde over the Fourier coefficients has been omitted.}:
\begin{equation}\label{pai11}
  \frac{\alpha_0^n}{2}+\sum_{m=1}^\infty\alpha_m^n=1\quad {\rm and} \quad \eta={\mathbb C}^{-1/n}
\end{equation}
\noindent with $\alpha_0$ and $\alpha_m$ denoting the Fourier coefficients of $f(x)$. So, the main function reads $f(x)={\mathbb C}^{-1/n}\phi(x)$. We define ($\alpha_j^{(0)},\alpha_{j,m})\equiv(|y|^{1/n}a_j\alpha_0,|y|^{1/n} a_j\alpha_m$) with $j=1,\cdots,n$.

  \subsection{Tasks of the Users}

  \noindent {\bf 1)} Users split their secret codes in four parts: $x_ja_j=a_j^{(1)}+a_j^{(2)}+a_j^{(3)}+a_j^{(4)}$;

  \noindent {\bf 2)} Users chose the masks $\omega_{j,m}=a_{j,m}+\Theta^{[1]}b_{j,m}$; ${\widehat\omega}_{j,m}=\Theta^{[1]}a_{j,m}-b_{j,m}$; $\omega_j^{(0)}=a_j^{(0)}+\Theta^{[1]}b_j^{(0)}$ and ${\widehat\omega}_j^{(0)}=\Theta^{[1]}a_j^{(0)}-b_j^{(0)}$ with $j=1,\cdots,n$. $\omega_j^{(0)}$, $\omega_j$, ${\widehat\omega}_j^{(0)}$ and ${\widehat\omega}_j$ are {\it numbers arbitrarily chosen by the users}. Note that is convenient to choose $\omega^{(0)}_{j}=\alpha^{(0)}(a^{(0)}_j+\Theta^{[1]}b^{(0)}_j)$, ${\widehat\omega}^{(0)}_{j}=\alpha^{(0)}(\Theta^{[1]}a^{(0)}_j-b^{(0)}_j)$, $\omega_{j,m}=(a_j+\Theta^{[1]}b_j)\alpha_m$, and ${\widehat\omega}_{j,m}=(\Theta^{[1]}a_j-b_j)\alpha_m$.

  \noindent {\bf 3)} Users construct the four {\it hyper-vectors} $A_j^{(1)}, A_j^{(2)}, A_j^{(3)}, A_j^{(4)}$ defined as
  \begin{eqnarray}\label{nU1}
    \!\!\!\!\!\!\!\!\!\!\!\!&&A_j^{(1)}\equiv\left\{a_j^{(1)},\alpha_j^{(0)}\!+\!\omega_j^{(0)},\alpha_{j,m}\!+\!\omega_{j,m}\right\} \nonumber \\
    \!\!\!\!\!\!\!\!\!\!\!\!&& A_j^{(3)}\equiv\left\{a_j^{(3)},\alpha_j^{(0)}\!+\!{\widehat\omega}_j^{(0)},\alpha_{j,m}\!+\!{\widehat\omega}_{j,m}\right\} \nonumber\\
    \!\!\!\!\!\!\!\!\!\!\!\!&&A_j^{(2)}\equiv\left\{a_j^{(2)},\alpha_j^{(0)}\!-\!\omega_j^{(0)},\alpha_{j,m}\!-\!\omega_{j,m}\right\} \nonumber \\
    \!\!\!\!\!\!\!\!\!\!\!\!&& A_j^{(4)}\equiv\left\{a_j^{(4)},\alpha_j^{(0)}\!-\!{\widehat\omega}_j^{(0)},\alpha_{j,m}\!-\!{\widehat\omega}_{j,m}\right\} \nonumber
  \end{eqnarray}

  \noindent {\bf 4)} Users send the hyper-vectors $A_j^{(1)}$ and $A_j^{(2)}$ to the {\it Node 1} and the {\it Node 2}, respectively, and the hyper-vectors $A_j^{3}$ and $A_j^{4}$ to the {\it Node 3} and {\it Node 4}, respectively.

  \subsection{Tasks of the Nodes}

  The four nodes perform the following tasks. Note that the sign is $+$ if $y>0$ and $-$ if $y<0$:

  \noindent {\it Node 1 computes}:
  \begin{dmath}\label{nU2}
    N_1\equiv \sum_{j=1}^na_j^{(1)}\pm\left(\frac{1}{8}{\widehat\Pi}_{j=1}^n(\alpha_j^{(0)}+\omega_j^{(0)})+\frac{1}{4}\sum_{m=1}^\infty{\widehat\Pi}_{j=1}^n(\alpha_{j,m}+\omega_{j,m})\right)\nonumber
  \end{dmath}

  \noindent {\it Node 2 computes}:
  \begin{dmath}\label{nU3}
    N_2\equiv \sum_{j=1}^na_j^{(2)}\pm\left(\frac{1}{8}{\widehat\Pi}_{j=1}^n(\alpha_j^{(0)}-\omega_j^{(0)})+\frac{1}{4}\sum_{m=1}^\infty{\widehat\Pi}_{j=1}^n(\alpha_{j,m}-\omega_{j,m})\right)\nonumber
  \end{dmath}

  \noindent {\it Node 3 computes}:
  \begin{dmath}\label{nU4}
    N_3\equiv \sum_{j=1}^na_j^{(3)}\pm\left(\frac{1}{8}{\widehat\Pi}_{j=1}^n(\alpha_j^{(0)}+{\hat\omega}_j^{(0)})+\frac{1}{4}\sum_{m=1}^\infty{\widehat\Pi}_{j=1}^n(\alpha_{j,m}+{\hat\omega}_{j,m})\right)\nonumber
  \end{dmath}

  \noindent {\it Node 4 computes}:
  \begin{dmath}\label{nU5}
    N_4\equiv \sum_{j=1}^na_j^{(4)}\pm\left(\frac{1}{8}{\widehat\Pi}_{j=1}^n(\alpha_j^{(0)}-{\hat\omega}_j^{(0)})+\frac{1}{4}\sum_{m=1}^\infty{\widehat\Pi}_{j=1}^n(\alpha_{j,m}-{\hat\omega}_{j,m})\right)\nonumber
  \end{dmath}

  \subsection{Task of the Display}

\begin{figure*}[htb]
    \centering
    \includegraphics[width=10cm,height=7cm]{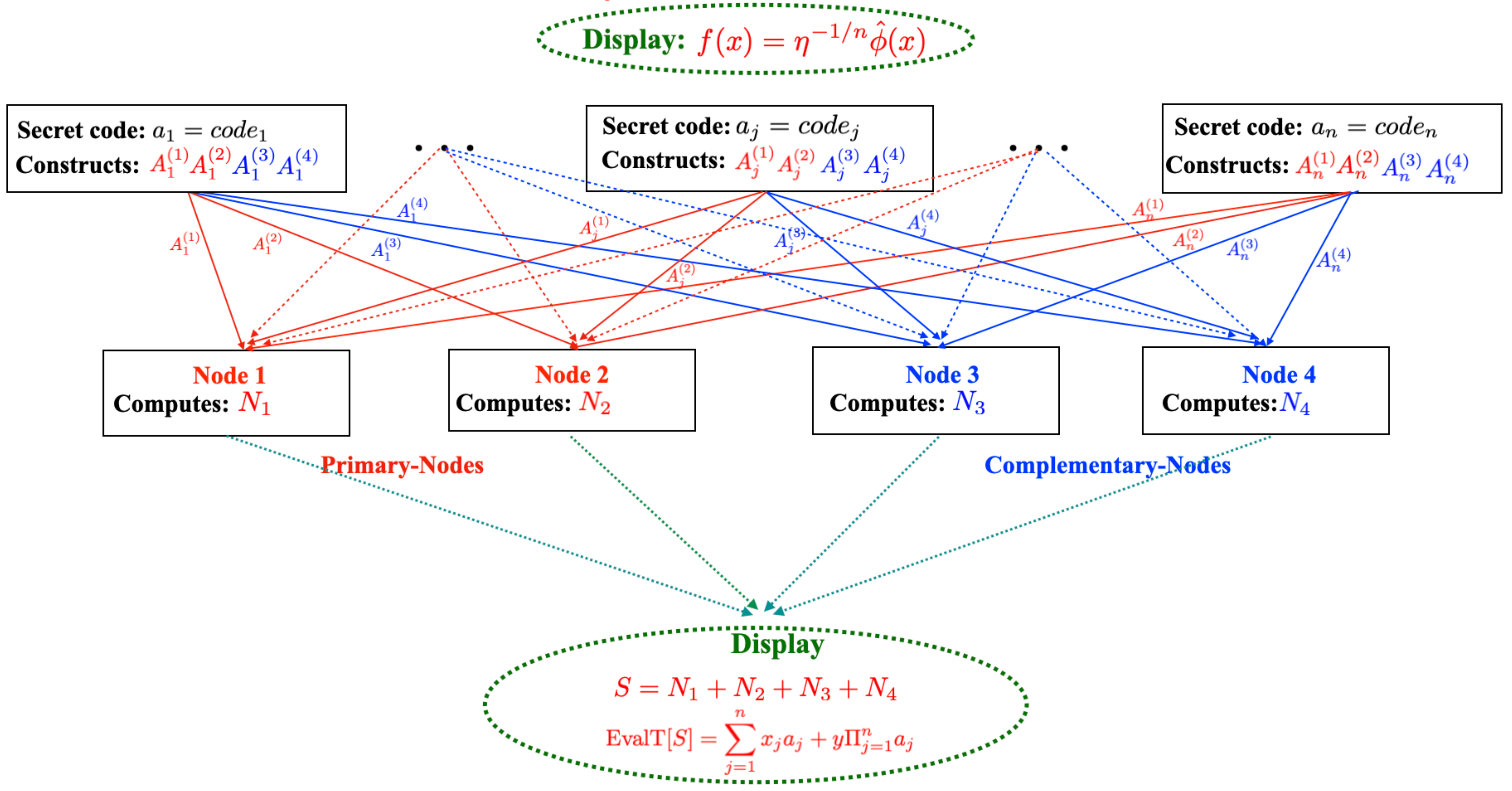}
    \caption{
      Display of the expression $\sum_{j=1}^nx_ja_j+y\Pi_{j=1}^na_j$ by keeping secret the codes $a_j$.
      \label{fig_nU}}
  \end{figure*}

  \noindent The display computes:
  \begin{equation}\label{nU6}
    S=N_1+N_2+N_3+N_4\nonumber
  \end{equation}
  \noindent The display shows:
  \begin{equation}\label{nU7}
    {\rm EvalT}[S]=\sum_{j=1}^nx_ja_j+y\Pi_{j=1}^na_j\nonumber
  \end{equation}
  \noindent due to the {\it generalized Parseval's identity} (\ref{pai11}). Note that to perform the operations, the Display takes into account the identities ${\rm EvalT}[\Theta^{[2m]}]=-1$ and ${\rm EvalT}[\Theta^{[2m-1]}]=+i$. The privacy of the participants is guaranteed if at least three nodes are not corrupted. Figure~\ref{fig_nU} shows the procedure.

  \section{Display of the expression $\sum_{j=1}^nx_ja_j+y\Pi_{j=1}^na_j$ by using $N = 3f+1$ nodes}\label{nUN}

  At this stage, we have solved the problem for $n$ users, using $4$ nodes. Participants' privacy is guaranteed that at least three nodes are not corrupted. However, requiring that three out of four nodes must not be corrupted is a quite restrictive constraint. This drawback can easily be overcome by using multiple nodes, say $N=3f + 1$ with $f$ denoting a natural number. The problem is solved if we can establish an algorithm that guarantees participants' privacy without modifying the constraint on the number of nodes that must not be corrupted. In other words, the number of nodes that are not to be corrupted must be $3$ and must not depend on $f$. We shall solve this problem using $3f + 1$ nodes, grouped in different $4$ categories, with the constraint that at least three nodes (and no more than three) belonging to three different categories must not be corrupted and at list one node, belonging to the {\it second level} of computation, is not corrupted. Firstly, it is convenient to account for all the quantities entered in the algorithm by a unique index, say $\iota$, that can take only natural numbers (i.e., $\iota = 0,1,2, \cdots$). We group the $3f+1$ nodes  in $4$ different categories. Let us denote with $\kappa$ the number of nodes in each category. If the total nodes is $N=3f+1$, we have $\kappa=3(f-1)/4$. Hence, in terms of $\iota$:
  \begin{eqnarray}\label{nUN1}
    && f=1+4\iota\nonumber\\
    && \kappa=3\iota\nonumber\\
    &&N=4(1+3\iota)\quad {\rm with}\quad \iota=0,1,2,\cdots \nonumber
  \end{eqnarray}

  \subsection{Tasks of the Users}

  \noindent {\bf 1)} Users split the codes in $12\iota$ parts: $x_ja_j=a_j^{(1)}+a_j^{(2)}+\cdots+a_j^{(12\iota)}$;

  \noindent {\bf 2)} Each user ($j$) chooses:

  \noindent {\bf 2a)} $\kappa$ {\it positive supplementary masks}: $\lambda_j^{(0,s)}$, $\mu_j^{(0,s)}$, $\nu_j^{(0,s)}$, and $\sigma_j^{(0,s)}$ with $s=1,\cdots,\kappa$;

  \noindent {\bf 2b)} $\kappa$ positive supplementary {\it multiplicative masks}: $\lambda_j^{(s)}$, $\mu_j^{(s)}$, $\nu_j^{(s)}$, and $\sigma_j^{(s)}$ with $s=1,\cdots,\kappa$;

  \noindent {\bf 3)} Users form the {\it hyper-vectors} $A^{(1)}-A^{(\kappa)}$, $B^{(\kappa+1)}-B^{(2\kappa)}$, $C^{(2\kappa+1)}-C^{(3\kappa)}$, and $D^{(3\kappa+1)}-B^{(4\kappa)}$;

  \noindent {(\bf 4)} Users send the encrypted codes to the nodes.

  \subsection{Tasks of the Nodes in the Categories}

  \noindent The $N=4(1+3\iota)$ nodes are grouped in $4$ categories $A$, $B$, $C$, and $D$ built as follows:

  \noindent {\bf A = CATEGORY 1}

  \noindent Category A receives:
  \begin{eqnarray}\label{nUN2}
    &&A^{(1)}\equiv\Bigl\{\!S^{(1)}=\sum_{j=1}^na_j^{(1)}, \nonumber \\
    &&P\!A^{(0,1)}={\widehat\Pi}_{j=1}^n(\alpha_j^{(0)}\lambda_j^{(0,1)}+\!\omega_j^{(0)}\lambda_j^{(0,1)})^{1/\kappa}, \nonumber \\
    &&PA^{(1)}={\widehat\Pi}_{j=1}^n(\alpha_{j,m}\lambda_j^{(1)}+\omega_{j,m}\lambda_j^{(1)})^{1/\kappa}\Bigr\} \nonumber
  \end{eqnarray}

$\cdots$

  \begin{eqnarray}
    && A^{(\kappa-1)}\equiv\Bigl\{\!S^{(\kappa-1)}=\sum_{j=1}^na_j^{(\kappa-1)}, \nonumber \\
    && PA^{(0,\kappa-1)}={\widehat\Pi}_{j=1}^n(\alpha_j^{(0)}\lambda_j^{(0,\kappa-1)}+\omega_j^{(0)}\lambda_j^{(0,\kappa-1)})^{1/\kappa} \nonumber \\
    && PA^{(\kappa-1)}={\widehat\Pi}_{j=1}^n(\alpha_{j,m}\lambda_j^{(\kappa-1)}+\omega_{j,m}\lambda_j^{(\kappa-1)})^{1/\kappa}
    \Bigr\} \nonumber
    \nonumber
  \end{eqnarray}

  \begin{eqnarray}
    && A^{(\kappa)}\equiv\Bigl\{S^{(\kappa)}=\sum_{j=1}^na_j^{(\kappa)}, \nonumber \\
    && PA^{(0)}={\widehat\Pi}_{j=1}^n(\alpha_j^{(0)}{\widetilde\lambda}_j^{(0)}+\omega_j^{(0)}{\widetilde\lambda}_j^{(0)})^{1/\kappa},
    \nonumber\\
    && PA={\widehat\Pi}_{j=1}^n(\alpha_{j,m}{\widetilde\lambda}_j+\omega_{j,m}{\widetilde\lambda}_j)^{1/\kappa}\Bigr\}\nonumber
  \end{eqnarray}

  \noindent {\bf B = CATEGORY 2}

  \noindent Category B receives:

  \begin{eqnarray}\label{nUN3}
    && B^{(1)}\equiv\Bigl\{S^{(\kappa+1)}=\sum_{j=1}^na_j^{(\kappa+1)}, \nonumber \\
    && P\!B^{(0,1)}={\widehat\Pi}_{j=1}^n(\alpha_j^{(0)}\mu_j^{(0,1)}-\omega_j^{(0)}\mu_j^{(0,1)})^{1/\kappa}, \nonumber\\
    && PB^{(\kappa)}={\widehat\Pi}_{j=1}^n(\alpha_{j,m}\mu_j^{(1)}
    -\omega_{j,m}\mu_j^{(1)}
    )^{1/\kappa}\Bigr\} \nonumber
  \end{eqnarray}

$\cdots$

  \begin{eqnarray}
    && B^{(\kappa-1)}\equiv\!\Bigl\{S^{(2\kappa-1)}=\sum_{j=1}^na_j^{(2\kappa-1)}, \nonumber \\
    && PB^{(0,\kappa-1)}={\widehat\Pi}_{j=1}^n(\alpha_j^{(0)}\mu_j^{(0,\kappa-1)}-\omega_j^{(0)}\mu_j^{(0,\kappa-1)})^{1/\kappa},\nonumber\\
    && PB^{(\kappa-1)}={\widehat\Pi}_{j=1}^n(\alpha_{j,m}\mu_j^{(\kappa-1)}-\omega_{j,m}\mu_j^{(\kappa-1)})^{1/\kappa}\Bigr\} \nonumber
  \end{eqnarray}

  \begin{eqnarray}
    && B^{(\kappa)}\equiv\Bigl\{S^{(2\kappa)}=\sum_{j=1}^na_j^{(2\kappa)}, \nonumber \\
    && PB^{(0)}={\widehat\Pi}_{j=1}^n(\alpha_j^{(0)}{\widetilde\mu}_j^{(0)}-\omega_j^{(0)}{\widetilde\mu}_j^{(0)})^{1/\kappa},
    \nonumber\\
    && PB={\widehat\Pi}_{j=1}^n(\alpha_{j,m}{\widetilde\mu}_j-\omega_{j,m}{\widetilde\mu}_j)^{1/\kappa}\Bigr\} \nonumber
  \end{eqnarray}

  \noindent {\bf C = CATEGORY 3}
  \noindent Category C receives:

  \begin{eqnarray}\label{nUN4}
    && C^{(1)}\equiv\Bigl\{S^{(2\kappa+1)}=\sum_{j=1}^na_j^{(2\kappa)}, \nonumber \\
    && P\!C^{(0,1)}={\widehat\Pi}_{j=1}^n(\alpha_j^{(0)}\nu_j^{(0,1)}+{\widehat\omega}_j^{(0)}\nu_j^{(0,1)})^{1/\kappa},
    \nonumber\\
    && PC^{(1)}={\widehat\Pi}_{j=1}^n(\alpha_{j,m}\nu_j^{(1)}+{\widehat\omega}_{j,m}\nu_j^{(1)})^{1/\kappa}\Bigr\}  \nonumber
  \end{eqnarray}

$\cdots$

  \begin{eqnarray}
    && C^{(\kappa-1)}\equiv\Bigl\{\!S^{(3\kappa-1)}=\sum_{j=1}^n\!a_j^{(3\kappa-1)},\nonumber \\
    && PC^{(0,\kappa-1)}={\widehat\Pi}_{j=1}^n(\alpha_j^{(0)}\nu_j^{(0,\kappa-1)}+{\widehat\omega}_j^{(0)}\nu_j^{(0,\kappa-1)})^{1/\kappa}, \nonumber\\
    && PC^{(\kappa-1)}={\widehat\Pi}_{j=1}^n(\alpha_{j,m}\nu_j^{(\kappa-1)}+{\widehat\omega}_{j,m}\nu_j^{(\kappa-1)})^{1/\kappa}\Bigr\}
    \nonumber
  \end{eqnarray}

  \begin{eqnarray}
    && C^{(\kappa)}\equiv\Bigl\{S^{(3\kappa)}=\sum_{j=1}^n\!a_j^{(3\kappa)}, \nonumber \\
    && P\!C^{(0)}={\widehat\Pi}_{j=1}^n(\alpha_j^{(0)}{\widetilde\nu}_j^{(0)}+{\widehat\omega}_j^{(0)}{\widetilde\nu}_j^{(0)})^{1/\kappa}, \nonumber\\
    && PC={\widehat\Pi}_{j=1}^n(\alpha_{j,m}{\widetilde\nu}_j+{\widehat\omega}_{j,m}{\widetilde\nu}_j)^{1/\kappa}
    \Bigr\}
    \nonumber
  \end{eqnarray}

  \noindent {\bf D = CATEGORY 4}
  \noindent Category D receives:

  \begin{eqnarray}\label{nUN5}
    && D^{(1)}\equiv\Bigl\{S^{(3\kappa+1)}=\sum_{j=1}^na_j^{(3\kappa)},  \nonumber\\
    && PD^{(0,1)}={\widehat\Pi}_{j=1}^n(\alpha_j^{(0)}\sigma_j^{(0,1)}-{\widehat\omega}_j^{(0)}\sigma_j^{(0,1)})^{1/\kappa}, \nonumber\\
    && PD^{(1)}={\widehat\Pi}_{j=1}^n\alpha_{j,m}(\sigma_j^{(1)}-{\widehat\omega}_{j,m}\sigma_j^{(1)})^{1/\kappa}\Bigr\}
    \nonumber
  \end{eqnarray}

$\cdots$

  \begin{eqnarray}
    && D^{(\kappa-1)}\equiv\Bigl\{S^{(4\kappa-1)}=\sum_{j=1}^na_j^{(4\kappa-1)}, \nonumber \\
    && PD^{(0,\kappa-1)}={\widehat\Pi}_{j=1}^n(\alpha_j^{(0)}\sigma_j^{(0,\kappa-1)}-{\widehat\omega}_j^{(0)}\sigma_j^{(0,\kappa-1)})^{1/\kappa},
    \nonumber\\
    && PD^{(\kappa-1)}={\widehat\Pi}_{j=1}^n(\alpha_{j,m}\sigma_j^{(\kappa-1)}-{\widehat\omega}_{j,m}\sigma_j^{(\kappa-1)})^{1/\kappa}\Bigr\}
    \nonumber
  \end{eqnarray}

  \begin{eqnarray}
    && D^{(\kappa)}\equiv\Bigl\{S^{(4\kappa)}=\sum_{j=1}^na_j^{(4\kappa)}, \nonumber \\
    && PD^{(0)}={\widehat\Pi}_{j=1}^n(\alpha_j^{(0)}{\widetilde\sigma}_j^{(0)}-{\widehat\omega}_j^{(0)}{\widetilde\sigma}_j^{(0)})^{1/\kappa}, \nonumber\\
    && PD={\widehat\Pi}_{j=1}^n(\alpha_{j,m}{\widetilde\sigma}_j-{\widehat\omega}_{j,m}{\widetilde\sigma}_j)^{1/\kappa}\Bigr\}
    \nonumber
  \end{eqnarray}

  \noindent Masks $\lambda_j$, $\mu_j$,$\nu_j$, and $\sigma_j$ $\lambda_j^{(0,s)}$, $\lambda_j^{(s)}$, etc. are {\it real numbers} different from zero and the variable with the tilde stands for the {\it inverse of the variable}, i.e.,
  \begin{eqnarray}\label{nUN6}
    &&\!\!\!\!\!\!\!\!\!\!\!\!\!\!\!\!\lambda_j{\widetilde\lambda}_j=1\quad ;\quad \mu_j{\widetilde\mu}_j=1\quad ; \quad \nu_j{\widetilde\nu}_j=1 \ ;\nonumber\\
    &&\!\!\!\!\!\!\!\!\!\!\!\!\!\!\!\!\sigma_j{\widetilde\sigma}_j=1\quad ;\quad {\lambda_j}^{(0,s)}{{\widetilde\lambda}_j}^{(0,s)}=1\quad
    {\rm etc.}\quad  (j=1,\cdots,n)\nonumber
  \end{eqnarray}
  \noindent Moreover,
  \begin{eqnarray}\label{nUN7}
    \!\!\!\!\!\!\!\!\!\!\!\!\!\!\!\!{\widetilde\lambda}_j^{(0)}\equiv\Pi_{s=1}^{\kappa-1}{\widetilde\lambda}_j^{(0,s)}\ ,\ {\widetilde\mu}_j^{(0)}\equiv\Pi_{s=1}^{\kappa-1}{\widetilde\mu}_j^{(0,s)}\  \nonumber\\
    {\widetilde\nu}_j^{(0)}\equiv\Pi_{s=1}^{\kappa-1}{\widetilde\nu}_j^{(0,s)} \ ,\ {\widetilde\sigma}_j^{(0)}\equiv\Pi_{s=1}^{\kappa-1}{\widetilde\sigma}_j^{(0,s)}\
    \nonumber\\
    \!\!\!\!\!\!\!\!\!\!\!\!\!\!\!\!
    {\widetilde\lambda}_j\equiv\Pi_{s=1}^{\kappa-1}{\widetilde\lambda}_j^{(s)}\ ,\ {\widetilde\mu}_j\equiv\Pi_{s=1}^{\kappa-1}{\widetilde\mu}_j^{(s)}\  \nonumber\\
    {\widetilde\nu}_j\equiv\Pi_{s=1}^{\kappa-1}{\widetilde\nu}_j^{(s)} \ ,\ {\widetilde\sigma}_j\equiv\Pi_{s=1}^{\kappa-1}{\widetilde\sigma}_j^{(s)}\qquad \forall j
    \nonumber
  \end{eqnarray}
  \subsection{Tasks of the Nodes $N_1$, $N_2$, $N_3$ and $N_4$}
  \noindent The nodes in the four categories perform the following tasks. Note that the sign is $+$ if $y>0$ and $-$ if $y<0$:

  \noindent $N_1$ computes:

  \noindent
  \begin{dmath}
    N_1 = \sum_{s=1}^\kappa S^{(s)}\pm\left(\frac{1}{8}PA^{(0)}\cdot \Pi_{s=1}^{\kappa-1}PA^{(0,s)}+{\frac{1}{4}PA\cdot \Pi_{s=1}^{\kappa-1} PA^{(s)}}\right)
  \end{dmath}

  \noindent $N_2$ computes:

  \begin{dmath}
    N_2 = \sum_{s=\kappa+1}^{2\kappa} S^{(s)}\pm\left(\frac{1}{8}PB^{(0)}\cdot \Pi_{s=1}^{\kappa-1}PB^{(0,s)}+\frac{1}{4}PB\cdot \Pi_{s=1}^{\kappa-1} PB^{(s)}\right)
  \end{dmath}

  \noindent $N_3$ computes:

  \begin{dmath}N_3 = \sum_{s=2\kappa+1}^{3\kappa} S^{(s)}\pm\left(\frac{1}{8}PC^{(0)}\cdot \Pi_{s=1}^{\kappa-1}PC^{(0,s)}+\frac{1}{4}PC\cdot \Pi_{s=1}^{\kappa-1} PC^{(s)}\right)  \end{dmath}

  \noindent $N_4$ computes:

  \begin{dmath}N_4 = \sum_{s=3\kappa+1}^{4\kappa} S^{(s)}\pm\left(\frac{1}{8}PD^{(0)}\cdot \Pi_{s=1}^{\kappa-1}PD^{(0,s)}+\frac{1}{4}PD\cdot \Pi_{s=1}^{\kappa-1} PD^{(s)}\right)  \end{dmath}

  \noindent To perform calculations, the nodes use rule {\bf vi)} of the $\Theta^{[n]}$-algebra, shown in Appendix~\ref{C}.
  \subsection{Task of the Display}
  \noindent The display computes: $S=N_1+N_2+N_3+N_4$ and shows publicly

\begin{figure*}[htb]
    \centering
    \includegraphics[width=13cm,height=9cm]{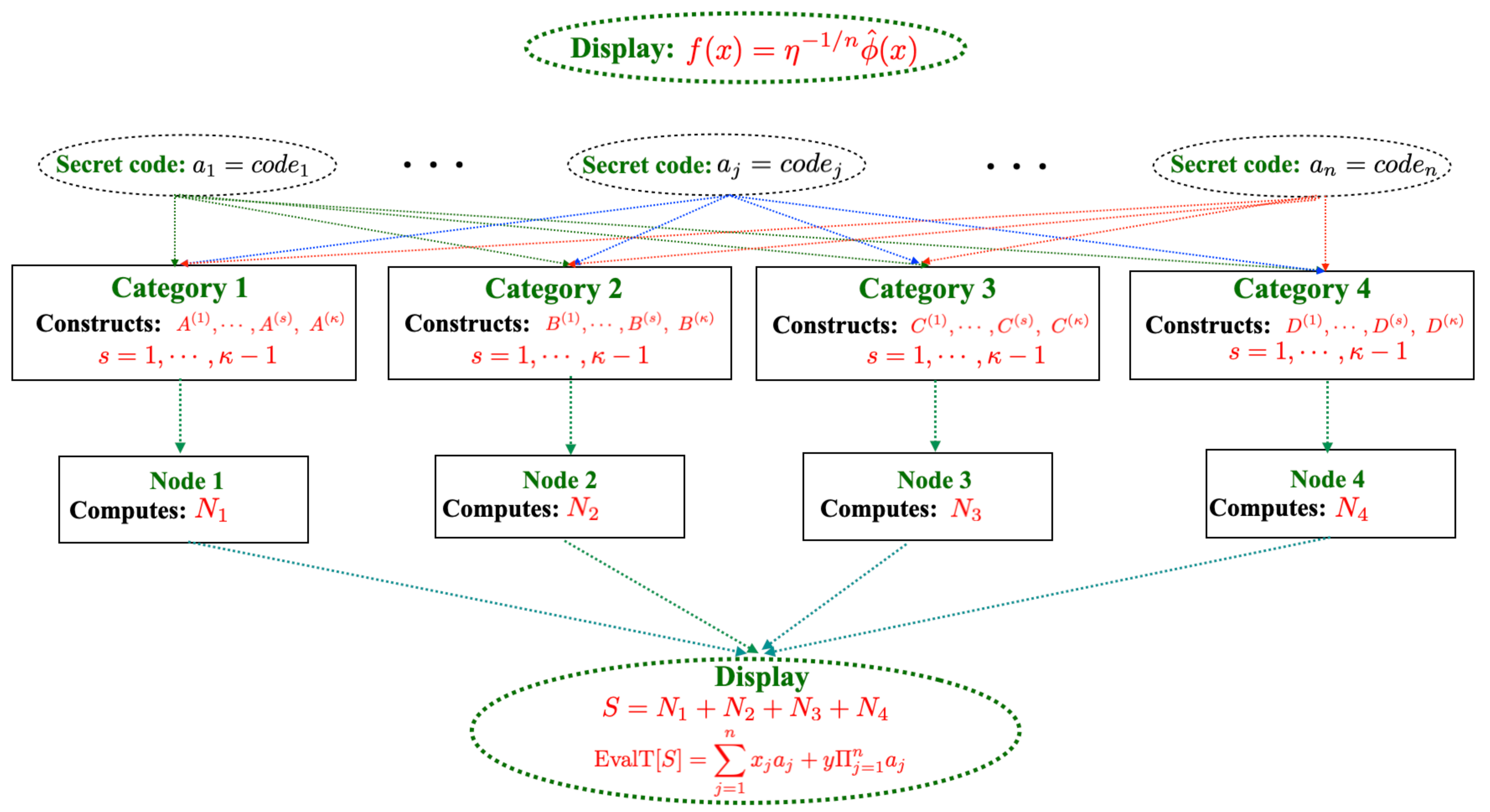}
    \caption{
      Display of the expression $\sum_{j=1}^nx_ja_j+y\Pi_{j=1}^na_j$ by using $N = 3f+1$ nodes
      \label{fig_nUN}}
  \end{figure*}

  \noindent ${\rm EvalT}[S]=\sum_{j=1}^nx_ja_j+y\Pi_{j=1}^na_j$.

  \noindent due to the {\it generalized Parseval's identity} (\ref{pai11}). The privacy of the participants is guaranteed if at least three nodes belonging to three different categories are not corrupted. Figure~\ref{fig_nUN} illustrates the procedure.

  \subsection{Considerations}
  To ensure the participants' privacy we have to impose the condition that at least three nodes, belonging to three different categories, are not corrupted. However, if we want to ensure that also the particular contribution of the addition or the multiplication to the final expression cannot be detected, we are bound to add that at least one of the four nodes $Node\ 1$, $Node\ 2$, $Node\ 3$ and $Node\ 4$ is not corrupted. Of course, in the second level of calculation, instead of 4 nodes we may add an arbitrary number of nodes without changing the restriction that at least one of these nodes must not be corrupted.

  \section{Practical Considerations}\label{cons}
  The designs described in the previous sections rely on nodes on the side to compute mathematical expressions. In this section, we present designs that incorporate our nodes within the infrastructure of a number of semi-permissioned Blockchains. This enables the execution of our protocol as a side effect of the normal system operations, taking no additional dependency on extra authorities. It remains an open problem how to embed our protocol into permissionless systems~\cite{bitcoin, ethereum}, based on proof of work or stake. These systems have a highly dynamic set of authorities maintaining the state of their blockchains, which cannot readily be mapped into the nodes of our system.

  Integration of our system into Hyperledger Fabric~\cite{hyperledger}---an example of a permissioned blockchain platform---is straightforward. Fabric contracts run on private sets of computation authorities---and use the Fabric protocols for cross-contract calls~\cite{ethereum}. In this setting, our computing nodes can coincide with the Fabric smart contract authorities. Upon a contract set up, they assign nodes into the different categories and then compute mathematical expressions when authorized by the contract. To compute these expressions, the nodes maintain a number of secret values (see \Cref{secret2Dim}) along with their traditional signing key needed for normal system operations. Integrating our system into blockchains has obvious advantages over traditional smart contracts capable of only evaluating expression over public inputs---as currently present in the Hyperledger. In other words, our system can augment the capabilities of any permissioned blockchain to enable computations over secret inputs. The threshold trust assumption---namely that integrity and availability are guaranteed under the corruption of a subset of authorities~\cite{hyperledger} is preserved, but need to consider the different categories of our nodes.

  We can also naturally embed our system into sharded scalable blockchains, as exemplified by Chainspace~\cite{chainspace}.
  In these systems, transactions are distributed and executed on `shards' of authorities, whose membership and public keys are known.
  Our computing nodes can naturally coincide with the authorities within a shard---a special object in Chainspace can signal to them to perform a private computation (rather than the default public computations). The authorities then attach their output to the transaction they are processing anyway. The trust assumptions of sharded blockchains naturally compose with the requirements of our computing nodes: each shard simply contains a node of every one of our four categories.

  \section{Related Works} \label{sec:related}
  There are two main constructions of multiparty protocols: circuit garbling and secret-sharing. Circuit garbling involves encrypting keys in a specific order to simulate a circuit evaluation~\cite{applebaum2014garble}; secret-sharing based protocol (as the one described in this paper) breaks the inputs among all nodes who use their shares to evaluate some function through local computations~\cite{bendlin2011semi,damgaard2013constant,nielsen2012new,lindell2015efficient}.

  SPDZ~\cite{SPDZ} is one of the most notorious secret-sharing based multiparty computation protocols scaling to an arbitrary number of users; SPDZ is secure against active adversaries using MACs to verify the integrity of computations, and does not require any kind of trusted third parties; it requires however expensive somewhat homomorphic encryption (SHE) to generate the triples used to compute multiplication of secrets. SPDZ2~\cite{SPDZ2} offers various improvements to the offline phase of SPDZ and allows the MACs to be checked without revealing their key, thus allowing the MAC to be re-used after it is checked. Mascot~\cite{mascot} uses oblivious transfer rather than SHE to further improve performances of the offline phase and generate triples.

  The literature following SPDZ mainly improves the offline phase, while our system innovates on both the offline and online phases.
  Most multiparty protocols for arithmetic circuits based on secret-sharing that scale to an arbitrary number of users is based on the algebra introduced by Donald Beaver~\cite{beaver1991efficient}. They thus require triples to compute the multiplication of secrets and impose communication between nodes during the online phase; their online latency, therefore, increases with the number of multiplications to evaluate. Our system comes with a different trade-off: our nodes do not communicate during the online phase and thus enjoy constant (and low) online latency in the size of the circuit, at the cost of not supporting the composition of operations (see \Cref{sec:limitations}). Established secret-sharing protocols face a trade-off between security and online latency---adding nodes improves security but increases latency. Our protocol forgoes this trade-off since multiplications do not require communication between the nodes.

  \section{Limitations and Future Work} \label{sec:limitations}
  Our system has several limitations that are beyond the scope of this work and deferred to future work.
  We do not support the composition of operations. That is, while most established scheme~\cite{SPDZ,SPDZ2,mascot} can evaluate expressions like $(a+b)(c+d)$ with two additions and one multiplication, we need to distribute the operation and evaluate $(ac+ad+bc+bd)$. We also defer future work adapting our scheme to withstand active adversaries, potentially adapting the MAC-based approach introduced by SPDZ~\cite{SPDZ}.

  \noindent As we have seen, as an example of a solution, for the case of $3f + 1$ nodes we have proposed a protocol that includes four categories with the addition of 4 nodes in the {\it second level} of calculation. We have imposed the condition that at least three nodes, belonging to three different categories, are not corrupted and that at least one of the nodes belonging to the {\it second level} of calculation is not corrupted. This is one of the ways to solve the problem, but we imagine that simpler and less restrictive protocols may be proposed. This too will be the subject of future work.

  \section{Discussion}\label{conc}
  This work aims to solve the following problem: {\it Propose a method able to show publicly a general mathematical expression while keeping the players' codes secret}. The problem has been solved by using the so-called {\it Masks' method}. In short, this method consists in hiding the codes of the players within the parameters of the functions and being able to show only the numerical value taken by the mathematical expression by using the {\it (generalized) Parseval identity}. The problem has been solved by using the method of {\it additive masks} and by applying the $\Theta^{[n]}$ algebra. Calculations are simplified by choosing an even main function $f(x)$. The codes remain secret if at least three out of four nodes are not corrupted. The values of a general mathematical expression are obtained by using interpolating polynomials. When a given function {\it Expr.} is replaced by Chebyshev interpolating polynomials, thanks to {\it theorem 1}, we can minimize the error, until reaching the "level of precision" allowed by the computers, using a sufficiently high number of polynomials. Moreover, {\it theorem 2} guarantees that Chebyshev method is the best among the other methods based on the 2nd degree of equations as its convergence rate is equal to 3. It is worth noting that the Chebyshev polynomials method is particularly convenient also in physical or engineering applications where the exact mathematical expression {\it Expr.} is not known, but only the (experimental) values that {\it Expr.} takes in the nodes are known.

  \noindent For treating simply the case of $3f+1$ {\it nodes}, we adopted the method of {\it multiplicative secondary masks} that, for simplicity, may be {\it arbitrary positive real numbers}. Even in this case, calculations are simplified by choosing an {\it even main function} $f(x)$. The nodes are organized in {\it four categories}, each of which contains $\kappa$ nodes. The nodes are labeled by the index $\iota$, which takes natural numbers $\iota=0,1,2,\cdots$. So, the numbers of nodes are $N=4+12\iota$, $f=1+4\iota$ and each category contains $\kappa=3\iota$ nodes. Hence, the total number of nodes is the nodes contained in the four categories plus $4$. The simplest case corresponds to $\iota=0$, i.e., the categories are empty. The participants' privacy is guaranteed if {\it at least three nodes belonging to three different categories are not corrupted} and one node belonging to the second level of calculation is not corrupted. In this case, no one can determine the codes of the players: the nodes are not able to determine the codes of the users and the users are not able to determine the codes of the other users.

  \section*{Data Availability Statement}

  \noindent The data that support the findings of this study are available from the corresponding author upon reasonable request.

  \section*{Acknowledgements}

  \noindent We thank George Danezis and Ioannis Psaras for helpful suggestions on the early manuscript and valuable advice.

  \bibliographystyle{alpha}
  \bibliography{references}

  \appendix

  \section{ Useful Relations for getting the Values of the Infinite Sums}\label{A}
  \begin{dmath}\label{A1}
    \sum_{m=1}^\infty\frac{(-1)^m}{m^2-\tau^2}=\frac{1}{2\tau^2}\left(1-\frac{\pi\tau}{\sin(\pi\tau)}\right)\ \ ;\ \ \sum_{m=1}^\infty\frac{1}{m^2-\tau^2}=\frac{1}{2\tau^2}\left(1-\pi\tau\cot(\pi\tau)\right)\nonumber
  \end{dmath}
  \noindent The sum of powers of these expressions may be obtained by performing the derivatives with respect to parameter $\tau$ e.g.,
  \begin{equation}\label{A2}
    \sum_{m=1}^\infty\frac{1}{(m^2-\tau^2)^2}=-\frac{1}{2\tau^4}+\frac{\pi\sin(2\pi\tau)+2\pi^2\tau}{8\tau^3\sin^2(\pi\tau)}\nonumber
  \end{equation}
  \noindent If, for example, we chose $f(x)=\eta^{-1/n}\cos({\pi\tau x/l})$ with $\eta={\mathbb C}^{-1}$, then
  \begin{equation}\label{A3}
    \alpha^{(0)}=2\eta^{-1/n}\frac{\sin(\pi\tau)}{\pi\tau}\quad{\rm and}\quad \alpha^{(m)}=\alpha^{(0)}\tau^2\frac{(-1)^m}{\tau^2-m^2}\nonumber
  \end{equation}
  \noindent For $n=2$ we get
  \begin{eqnarray}\label{A4}
    &&\eta=1+\frac{\sin(2\pi\tau)}{2\pi\tau} \nonumber \\
    && f(x)=\left(1+\frac{\sin(2\pi\tau)}{2\pi\tau}\right)^{-1/2}\!\!\!\!\!\!\!\!\cos(\pi\tau x/l)\nonumber\\
    &&\alpha^{(0)}=2\left(1+\frac{\sin(2\pi\tau)}{2\pi\tau}\right)^{-1/2}\frac{\sin(\pi\tau)}{\pi\tau} \nonumber \\
    && \alpha^{(m)}=\alpha^{(0)}\tau^2\frac{(-1)^m}{\tau^2-m^2}\nonumber
  \end{eqnarray}

  \section{Demonstration of the Generalized Parseval's Identity for Multiple Functions}\label{B}
  In this Section, we prove the validity of the generalized Parseval's identity given by Eq.~(\ref{pai8}). To this aim, let us first consider the constant $\mathbb C$ defined in Eq.~(\ref{pai5}) and the {\it cosine main functions} defined in Eq.~(\ref{pai1}). The convolution operation between two {\it cosine main functions} $\phi_c^{(\kappa_1)}(x)$ and $\phi_c^{(\kappa_2)}(x)$, with $\kappa_1$ and $\kappa_2$ integer numbers subject to the conditions $1\leq \kappa_1,\kappa_2\leq n\ ;\ \kappa_1\neq\kappa_2$, reads
  \begin{equation}\label{app1}
    (\phi_c^{(\kappa_1)}\star\phi_c^{(\kappa_2)})(x)=\frac{1}{2}\alpha_0^{(\kappa_1)}\alpha_0^{(\kappa_2)}+\sum_{m=1}^\infty\alpha_m^{(\kappa_1)}\alpha_m^{(\kappa_2)}\cos\Bigl(\frac{m\pi}{l}x\Bigr)\nonumber
  \end{equation}
  \noindent The convolution operation between this function and another cosine main function, say $\phi_c^{(\kappa_3)}(x)$ with $\kappa_3$ integer number subject to the conditions $1\leq\kappa_3\leq n\ ;\ \kappa_3\neq\kappa_1\neq\kappa_2$, gives
  \begin{dmath}\label{app2}
    \bigl((\phi_c^{(\kappa_1)}\star\phi_c^{(\kappa_2)})\star\phi_c^{(\kappa_3)}\bigr)(x)=\frac{1}{2}\alpha_0^{(\kappa_1)}\alpha_0^{(\kappa_2)}\alpha_0^{(\kappa_3)}+\sum_{m=1}^\infty\alpha_m^{(\kappa_1)}\alpha_m^{(\kappa_2)}\alpha_m^{(\kappa_3)}\cos\Bigl(\frac{m\pi}{l}x\Bigr)\nonumber
  \end{dmath}
  \noindent Hence, by performing $n-2$ convolution operations, sequentially, starting from $\phi_c^{(1)}(x)$ to $\phi_c^{(n-1)}(x)$ we get
  \begin{eqnarray}\label{app3}
    &&\bigl((\cdots(((\phi_c^{(1)}\star\phi_c^{(2)})\star\phi_c^{(3)})\star\phi_c^{(4)})\star\cdots)\star\phi_c^{(n-1)}\bigr)(x) \nonumber \\
    && \equiv Input1_C(x) \nonumber \\
    &&=\frac{1}{2}\Pi_{j=1}^{n-1}\alpha_0^{(j)}+\sum_{m=1}^{\infty}\Pi_{j=1}^{n-1}\alpha_m^{(j)}\cos\Bigl(\frac{m\pi}{l}x\Bigr)\nonumber
  \end{eqnarray}
  \noindent By integrating Eq.~(\ref{app3}) with $\phi_c^{(n)}(x)$, we finally have
  \begin{dmath}\label{app4}
    {\mathbb C}=\frac{1}{l}\int_{-l}^l\phi_c^{(n)}(x)Input1_C(x)dx=\frac{1}{2}\Pi_{j=1}^{n}\alpha_0^{(j)}+\sum_{m=1}^{\infty}\Pi_{j=1}^{n}\alpha_m^{(j)}
  \end{dmath}
  \noindent If we now perform the convolution operation between two {\it sine main functions} defined in Eq.~(\ref{pai1}), say $\phi_s^{(\kappa_1)}(x)$ and $\phi_s^{(\kappa_2)}(x)$ with $\kappa_1$ and $\kappa_2$ integer numbers subject to the conditions $1\leq \kappa_1,\kappa_2\leq n\ ;\ \kappa_1\neq\kappa_2$, we get
  \begin{equation}\label{app5}
    (\phi_s^{(\kappa_1)}\star\phi_s^{(\kappa_2)})(x)=-\sum_{m=1}^\infty\beta_m^{(\kappa_1)}\beta_m^{(\kappa_2)}\cos\Bigl(\frac{m\pi}{l}x\Bigr)\nonumber
  \end{equation}
  \noindent Hence, if $n$ is an even number, we may pair up two by two $n-2$ input functions and we may perform $n-3$ convolution operations between each pair by getting
  \begin{eqnarray}\label{app6}
    &&\bigl((\phi_s^{(1)}\star\phi_s^{(2)})\star(\phi_s^{(3)}\star\phi_s^{(4)})\star\cdots\star(\phi_s^{(n-3)}\star\phi_s^{(n-2)})\bigr)(x) \nonumber \\
    &&\equiv Input1_{S_e}(x)\nonumber\\
    &&Input1_{S_e}(x)=-\sum_{m=1}^\infty\Pi_{j=1}^{n-2}\beta_m^{(j)}\cos\Bigl(\frac{m\pi}{l}x\Bigr)\nonumber
  \end{eqnarray}
  \noindent Since
  \begin{equation}\label{app7}
    (\phi_s^{(n-1)}\star\phi_s^{(n)})(x)=-\sum_{m=1}^\infty\beta_m^{(n-1)}\beta_m^{(n)}\cos\Bigl(\frac{m\pi}{l}x\Bigr)\nonumber
  \end{equation}
  \noindent we finally obtain
  \begin{dmath}\label{app8}
    {\mathbb S}_e=\frac{1}{l}\int_{-l}^l(\phi_s^{(n-1)}\star\phi_s^{(n)})(x)Input1_{S_e}(x) dx=\sum_{m=1}^\infty\Pi_{j=1}^{n}\beta_m^{(j)}
  \end{dmath}
  \noindent Similarly, if $n$ is an odd number, firstly we note that
  \begin{equation}\label{app9}
    (\phi_s^{(n)}\wedge{\mathbb K})(x)=\sum_{m=1}^\infty\beta_m^{(n)}\cos\Bigl(\frac{m\pi}{l}x\Bigr)\nonumber
  \end{equation}
  \noindent Since
  \begin{eqnarray}\label{app10}
    &&\bigl((\phi_s^{(1)}\star\phi_s^{(2)})\star(\phi_s^{(3)}\star\phi_s^{(4)})\star\cdots\star(\phi_s^{(n-2)}\star\phi_s^{(n-1)})\bigr)(x) \nonumber \\
    && \equiv Input1_{S_o}(x)\nonumber\\
    &&=-\sum_{m=1}^\infty\Pi_{j=1}^{n-1}\beta_m^{(j)}\cos\Bigl(\frac{m\pi}{l}x\Bigr)\nonumber
  \end{eqnarray}
  \noindent we finally get
  \begin{equation}\label{app11}
    {\mathbb S}_o=-\frac{1}{l}\int_{-l}^l(\phi_s^{(n)}\wedge{\mathbb K})(x)Input1_{S_o}(x)dx=\sum_{m=1}^\infty\Pi_{j=1}^{n}\beta_m^{(j)}
  \end{equation}
  \noindent Adding Eq.~(\ref{app4}) to Eq.~(\ref{app8}) (if $n$ is an even number), or adding Eq.~(\ref{app4}) to Eq.~(\ref{app11}) (if $n$ is an odd number), we obtain the {\it generalised Parseval's identity} for $n$ inputs $\phi^{(j)}(x)$ (with $j=1,2,\cdots,n$)
  \begin{eqnarray}\label{app7}
    \!\!\!\!\!\!\!\!\!\!\!\!&&{\mathbb C}+{\mathbb S}_\kappa \\
    \!\!\!\!\!\!\!\!\!\!\!\! && =\frac{1}{2}\Pi_{j=1}^n\alpha_0^{(j)}+\sum_{m=1}^\infty\bigl(\Pi_{j=1}^n\alpha_m^{(j)}\bigr)+\sum_{m=1}^\infty\bigl(\Pi_{j=1}^n\beta_m^{(j)}\bigr)\ \ {\rm with} \nonumber \\
    \!\!\!\!\!\!\!\!\!\!\!\!&&{\mathbb S}_\kappa=
    \left\{ \begin{array}{ll}
      {\mathbb S}_e \  & \ \mbox{{\rm if}\ \ n={\rm even\ number}} \\
      {\mathbb S}_o\   & \ \mbox{{\rm if}\ \ n={\rm odd\ number}}  \\
    \end{array}
    \right.
    \nonumber
  \end{eqnarray}
  \noindent This theorem allows computing the normalization constant $\eta$:
  \begin{equation}\label{app13}
    \eta=({\mathbb C}+{\mathbb S}_\kappa)^{-1/n}\nonumber
  \end{equation}
  \noindent and the {\it normalised main functions} $f^{(j)}(x)$ read
  \begin{equation}\label{app14}
    f^{(j)}(x)=\eta\phi^{(j)}(x)\quad{\rm with}\quad j=1,2,\cdots, n\nonumber
  \end{equation}
  \noindent If we use the main functions $f^{(j)}(x)$ the generalised Parseval's identity reads
  \begin{equation}\label{app15}
    \frac{1}{2}\Pi_{j=1}^n{\widetilde\alpha}_0^{(j)}+\sum_{m=1}^\infty\bigl(\Pi_{j=1}^n{\widetilde\alpha}_m^{(j)}\bigr)+\sum_{m=1}^\infty\bigl(\Pi_{j=1}^n{\widetilde\beta}_m^{(j)}\bigr)=1
  \end{equation}
  \noindent with ${\widetilde\alpha}_0^{(j)}$, ${\widetilde\alpha}_m^{(j)}$, and ${\widetilde\beta}_m^{(j)}$ denoting the Fourier coefficients of $f^{(j)}(x)$. The schemes of calculation for coefficients ${\mathbb C}$, ${\mathbb S}_e$ and ${\mathbb S}_o$ are illustrated in Figure \ref{fig_C}, \ref{fig_Se}, and \ref{fig_So}, respectively.

  \section{Algebra of the symbol $\Theta^{[n]}$ and the Command EvalT[…]}\label{C}

  \noindent{\bf - $\Theta^{[n]}$ algebra}

  \noindent {\bf i)}
  \begin{eqnarray}
    && \Theta^{[1]}\ast\Theta^{[1]}\ast\Theta^{[1]}\ast\cdots\ast\Theta^{[1]}\equiv \Theta^{[n]}
    \quad (n\ {\rm times});\nonumber\\
    &&{\widehat\Pi}_{j=1}^nh_i\equiv h_1\ast h_2\ast\cdots\ast h_n
  \end{eqnarray}

  \noindent {\bf ii)}
  \begin{eqnarray}
    &&\Theta^{[n]}\ast 1=1\ast\Theta^{[n]}=\Theta^{[n]}; \nonumber\\
    &&\Theta^{[n]}\ast 0=0\ast\Theta^{[n]}=0
  \end{eqnarray}

  \noindent {\bf iii)}
  \begin{dmath}
    \Theta^{[m]}\ast(\Theta^{[n]}\ast\Theta^{[\kappa]})=(\Theta^{[m]}\ast\Theta^{[n]})\ast\Theta^{[\kappa]}=\Theta^{[m]}\ast\Theta^{[n]}\ast\Theta^{[\kappa]}=\Theta^{[m+n+\kappa]}
  \end{dmath}
  with $m$, $n$ and $k$ denoting {\it positive natural numbers}.

  \noindent {\bf iv)}
  \begin{dmath}
    \Theta^{[m]}\ast\Theta^{[n]}=\Theta^{[n]}\ast\Theta^{[m]}=\Theta^{[m+n]}$; $(\Theta^{[n]})^m\equiv\Theta^{[nm]}
  \end{dmath}

  \noindent {\bf v)}
  \begin{eqnarray}
    \!\!\!\!\!\!\!\!\!\!\!\!\!\!\!\!&&\Theta^{[n]}\ast(x+y)=\Theta^{[n]}\ast x+\Theta^{[n]}\ast y\nonumber\\
    \!\!\!\!\!\!\!\!\!\!\!\!\!\!\!\!&& = \Theta^{[n]}x+\Theta^{[n]} y=x\Theta^{[n]}+y\Theta^{[n]}\quad (x\ast y\equiv xy)
  \end{eqnarray}
  \noindent with $x$ and $y$ denoting {\it two complex numbers}.

  \noindent {\bf vi)}
  \begin{dmath}
    \Big(\big(x_1+x_2\Theta^{[n]}\big)^{1/\kappa}\Big)^\kappa\equiv x_1+x_2\Theta^{[n]}
  \end{dmath}

  \noindent with $x_1$, $x_2$ denoting {\it numbers} and $\kappa$ is a {\it real number}, respectively.
  \vskip0.2truecm
  \noindent {\bf - The command {\bf EvalT[…]}}

  \noindent {\bf vii)}
  \begin{dmath}
    {\rm EvalT}[\Theta^{[n]}]\equiv \exp\left(i\frac{\pi}{4}\left(3+(-1)^n\right)\right)
  \end{dmath}

  \noindent with $n$ denoting {\it positive natural numbers}.

  \noindent {\bf viii)}
  \begin{dmath}({\rm EvalT}[\Theta^{[n]}])^m\equiv \exp\left(i\frac{m\pi}{4}\left(3+(-1)^n\right)\right)
  \end{dmath}

  \noindent with $m$ denoting a {\it number}

  \noindent {\bf ix)}
  \begin{dmath}
    {\rm EvalT}[f({\bf x}_1,\Theta^{[n]})+g({\bf x}_2,\Theta^{[n]})]\equiv f({\bf x}_1,{\rm EvalT}[\Theta^{[n]}])+g({\bf x}_2,{\rm EvalT}[\Theta^{[n]}])
  \end{dmath} with $x_1$ and $x_2$ denoting {\it numbers}.

  \noindent {\bf x)}
  \begin{dmath}{\rm EvalT}[f({\bf x}_1,\Theta^{[n]}) g({\bf x}_2,\Theta^{[n]})]\equiv f({\bf x}_1,{\rm EvalT}[\Theta^{[n]}])g({\bf x}_2,{\rm EvalT}[\Theta^{[n]}])
  \end{dmath} with $f$ and $g$ denoting {\it general functions}

  \noindent with $x$, $y$ denoting {\it numbers}.

  \noindent {\bf xi)}
  \begin{dmath}
    {\rm EvalT}[f({\bf x},(\Theta^{[n]})^y)]\equiv f({\bf x},{\rm EvalT}[\Theta^{[ny]}]\end{dmath}

  \noindent The basic rule is to perform the operation $\ast$ first and then apply the command {\bf EvalT [...]} by taking into account the properties {\bf i)} - {\bf x)} above.
  \vskip 0.2truecm
  \noindent {\bf - Identities and Differences}

$({\rm EvalT}[\Theta^{[2n-1]}])^2={\rm EvalT}[\Theta^{[2m]}]=-1$

  \noindent with $m$ and $n$ denoting {\it integer numbers}.

  \begin{dmath}
    {\rm EvalT}[\Theta^{[n]}]{\rm EvalT}[\Theta^{[m]}]\neq{\rm EvalT}[\Theta^{[n+m]}]
  \end{dmath}
  \begin{dmath}
    ({\rm EvalT}[\Theta^{[n]}])^y\neq{\rm EvalT}[\Theta^{[ny]}]
  \end{dmath}

  \begin{dmath}
    {\rm EvalT}[f({\bf x}_1,\Theta^{[n]})\ast g({\bf x}_2,\Theta^{[n]})]\neq {\rm EvalT}[f({\bf x}_1,\Theta^{[n]})]{\rm EvalT}[g({\bf x}_2,\Theta^{[n]})]
  \end{dmath}

  \vskip 0.2truecm
  \noindent {\bf - Examples}

  \begin{dmath}
    (x_1+i\Theta^{[1]}y_1)\ast (x_2+i\Theta^{[2]}y_2)=x_1x_2+i\Theta^{[1]}x_2y_1+i\Theta^{[2]}x_1y_2-\Theta^{[3]}y_1y_2
  \end{dmath} with $x_1$, $x_2$, $y_1$, and $y_2$ denoting {\it numbers}.

\begin{dmath}
  {\rm EvalT}[(x_1+i\Theta^{[1]}y_1)\ast (x_2+i\Theta^{[2]}y_2)]={\rm EvalT}[x_1x_2+i\Theta^{[1]}x_2y_1+i\Theta^{[2]}x_1y_2-\Theta^{[3]}y_1y_2]= x_1x_2-x_2y_1-i(x_1y_2+y_1y_2)
\end{dmath}

\end{document}